\documentclass[manuscript]{aastex}
\usepackage[top=1in, bottom=1in, left=1.0in, right=1in]{geometry}
\usepackage{graphics,natbib,lscape}
\usepackage{multirow}
\usepackage{longtable}

\shorttitle{Binary Frequencies in Globular Clusters}
\shortauthors{Ji \& Bregman}

\begin{document}

\title{Binary Frequencies in a Sample of Globular Clusters. I. Methodology and Initial Results}

\author{Jun Ji and Joel N. Bregman}

\affil{Department of Astronomy, University of Michigan, Ann Arbor, MI 48109}
\email{jijun@umich.edu, jbregman@umich.edu}

\begin{abstract}

Binary stars are thought to be a controlling factor in globular cluster evolution, since they can heat the environmental stars by converting their binding energy to kinetic energy during dynamical interactions.
Through such interaction, the binaries determine the time until core collapse.  To test predictions of this model, we have determined binary fractions for 35 clusters.
Here we present our methodology with a representative globular cluster NGC 4590.
We use HST archival ACS data in the F606W and F814W bands and apply PSF-fitting photometry to obtain high quality color-magnitude diagrams.
We formulate the star superposition effect as a Poisson probability distribution function, with parameters optimized through Monte-Carlo simulations.
A model-independent binary fraction of (6.2$\pm0.3$)\% is obtained by counting stars that extend to the red side of the residual color distribution after accounting for the photometric errors and the star superposition effect.
A model-dependent binary fraction is obtained by constructing models with a known binary fraction and an assumed binary mass-ratio distribution function.
This leads to a binary fraction range of 6.8\% to 10.8\%, depending on the assumed shape to the binary mass ratio distribution, with the best fit occurring for a binary distribution that favors low mass ratios (and higher binary fractions).
We also represent the method for radial analysis of the binary fraction in the representative case of NGC 6981, which
shows a decreasing trend for the binary fraction towards the outside, consistent with theoretical predictions for the dynamical effect on the binary fraction.
\end{abstract}


\section{Introduction}
The standard picture of globular clusters shows that a cluster composed of single stars will undergo core collapse after several relaxation times \citep{lynden68,cohn80,lynden80,spitzer87}. Since only about one-fifth of globular clusters show collapsed cores \citep{Djorgovski86,harris96}, certain heating mechanisms are needed to counteract the gravitational contraction and avoid core collapse. This energy is expected to come primarily from the "burning" of binaries, i.e. the dynamical interactions of binaries with single stars or other binaries will convert the binding energy in the binaries to the single stars or other binaries, so as to heat the environment stars \citep{Heggie75,hut83,goodman89,hut92a}. Even a small primordial binary fraction is sufficient to prevent core collapse for many relaxation times \citep{gao91,freg03}, so the binary fraction is an essential parameter that can dramatically affect the evolution of globular clusters.

The binaries remaining in globular clusters are mainly hard binaries, whose binding energy is greater than the average kinetic energy of a single star in that cluster \citep{hut92a}. Most of the soft ones are destroyed during their first interactions with other stars \citep{sollima08}, and would not provide the heating energy. Mass-transfer binaries are among the hardest binaries, and those with degenerate primaries can be bright X-ray sources \citep{hut92a,Heinke11}, such as Low mass X-ray binaries (LMXBs). LMXBs are thought to be formed in the dense cores of globular clusters through dynamical exchange processes \citep{Clark75,Bailyn95,Cohn10}, as they show a strong correlation between the collisional parameter and their frequency \citep{Pooley03}. Some CVs show this correlation too, which suggests their dynamical origin \citep{Pooley06}.

Although theoretical models and simulations are well developed for the evolution of globular clusters with binary burning process, sufficient observations to test these models are lacking. One reason is that it is very difficult to isolate individual stars in the high density region such as the cores of globular clusters from the ground based telescope. The other reason is that photometry errors are very large due to superposition of those unresolved stars. These two difficulties make the observations of binaries fraction challenging.

A direct method to detect binaries in globular clusters is by spectroscopic observation to measure radial velocity variations, which can only be applied to red giant and sub-giant stars. This is because those stars are bright in magnitude and cool in temperature, so there are many strong absorption lines for cross-correlation. This will improve the accuracy of radial velocity measurement to below 1 km/s. The drawback for this method is that it requires large amounts of observing time over a several years.  For some long period binaries (greater than 10 years), even the current observational accuracy is not enough to discover the small radial velocity change over a reasonable time (5 years).
For binaries composed of two main sequence stars, radial velocity observation present other challenges.
Not only are these stars fainter than giants, but they have fewer lines for spectral cross-correlations, thereby demanding high S/N spectra that can place unreasonable demands on even the largest ground-based telescopes.
Another issue with ground-based spectroscopic observations is that, even for good seeing conditions, multiple unrelated stars can fall in a single spectral PSF.  This restricts such studies to the outer regions of globular clusters, which may not be representative of the dynamically more active inner parts \citep{Gunn79,Pryor88,Yan96a,cote96a,cote96b}.

A different direct method for detecting binaries is through observation of eclipsing binaries. This method still needs a substantial observing time investment to monitor many stars in globular clusters and search for the photometric variables through their light curves. While this method is biased by small orbital inclination short period binaries, it is another valuable method for investigating binary systems \citep{Mateo90,yan94,Yan96b,albrow01}.

Another method, and the one used here, makes use of the accurate measurement of stars that form the main sequence in a color-magnitude diagram (CMD).  For main sequence stars of a single age and metallicity, the width of the main sequence of single stars should be limited only by measurement error.
Relative to the single star main sequence, a binary star will be either redder, for a lower mass secondary, or brighter (by 0.75 mag) but at the same color for two stars of equal mass.
An ensemble of binaries forms a thickening to the red of the single star main sequence, which we can measure.
The method is difficult to execute from the ground because of the precision required at faint magnitudes (S/N = 50-100 at V = 21) and in crowded fields.
However, the Hubble Space Telescope (HST) has sufficient sensitivity and spatial resolution to resolve even the core of globular clusters. This greatly improves the photometric accuracy, and makes the measurement of the binary fraction possible as a function of position.
This method can detect widely-separated binaries, but it is biased against binaries with low-mass secondaries. The benefit of this method is that, it requires relatively less observing time, and one can measure the global binary fraction without other assumptions.

To date, only a few globular clusters have been studied for the binary fraction with HST by analysing their color-magnitude diagrams. \citet{rub97} measured that the binary fraction of NGC 6752 to be 15\% - 38\% in the inner core radius, and probably less than 16\% beyond that.
\citet{bell02} also determined the binary fraction of NGC 288 within its half light radius to be  $(15\pm5)\%$ with HST WFPC2 data, depending on the adopted binary mass ratio function. Its binary fraction outside the half light radius is less than 10\%, and most likely closer to 0\%.
\citet{rich04} measured the evolved cluster M4 with proper-motion selected WFPC2 data. They found the binary fraction within 1.5 core radius is about 2\%, decreasing to about 1\% between 1.5 and 8.0 core radii.  \citet{zhao05} studied two clusters, M3 and M13, with WFPC2 data, and they found that the binary fraction of M3 within one core radius lies between 6\% and 22\%, and falls to 1\% - 3\% between 1-2 core radii. The binary fraction of M13, however, was not constrained with their method.
\citet{davis08} measured the binary fraction of the core collapsed cluster NGC 6397 with both ACS and WFPC2 data. They well constrained the binary fraction outside the half light radius to be $(1.2\pm0.4)\%$ with 126 orbits of observation with ACS and proper-motion selected clean data, and constrained the binary fraction within the half light radius to be $(5\pm1)\%$ with WFPC2 data.

\citet{sollima07} performed the first sample study for the binary fractions in globular clusters. They used aperture photometry to construct the CMDs for 13 low-density, high galactic latitude globular clusters with ACS data. They found a minimum of 6\% binary fraction within one core radius for all clusters in their sample, and global fractions ranging from 10 to 50 per cent depending on the clusters and the assumed binary mass-ratio model.

In our survey, we compile a sample of 35 Galactic globular clusters, a larger sample than previous efforts, and one that takes advantage of a wealth of archived HST data.  We use the PSF photometry instead of aperture photometry, so that we can analyze high-density clusters in addition to low-density clusters.  We try to constrain the binary mass-ratio function depending on the data quality and the number of binaries found, which is the largest uncertainty in determining binary fraction in previous studies. We also analyze the binary fraction radial distribution and variation along the main-sequence.  In this paper, the first of two, we present the techniques used in obtaining the binary fraction. We present the sample selection in $\S$\ref{sample}, data reduction and photometry method in
 $\S$\ref{sec:reduction}, the artificial star tests in $\S$\ref{sec:artificial}, the high mass-ratio binary fraction estimate in $\S$\ref{sec:high-q fb}, the global binary fraction estimate in $\S$\ref{sec:global Fb}, the binary fraction radial analysis in $\S$\ref{sec:radial}, and discussions in $\S$\ref{sec:discussion}.

\section{Sample Selection}
\label{sample}
To achieve the primary goal of determining the binary fraction, we need accurate
color-magnitude diagrams (CMD).  CMDs can be made from several different filters, such as a subset of
B, V, R, and I, although for the main sequence in the late G to early M spectral region, the most accurate CMDs
are composed from a subset of V, R, and I (V and I to be used when possible).

For good photometric errors (0.02 mag rms), about
6,000 stars on the CMD will yield an uncertainty in the binary fraction of 1.5-2\% (after \citet{hut92a},
Table 4).  This type of photometric accuracy is achieved with WFPC2 V band images in 1500
seconds for V = 24.0 (m-M = 14.6 mag for a M1 star), or for V = 24.9 when using
the ASC/WFC (m-M = 15.5 mag for a M1 star).

We examined the HST observations taken for every Galactic globular cluster in the list
of  \citet{harris96} and most have some WFPC2 or ACS observations, but many of these were taken
in snapshot mode and are not sufficiently long to meet our criteria.  We find about 35 globular
clusters that are sufficiently  luminous, and with long enough
exposures that we can extract useful information on binary fractions. Most of those data sets are from
the HST Treasury program for globular clusters with ACS observations in F606W and F814W filters \citep{Saraj07}.

\begin{figure}[htbp] \centering
\includegraphics[width=2.75in,angle=0]{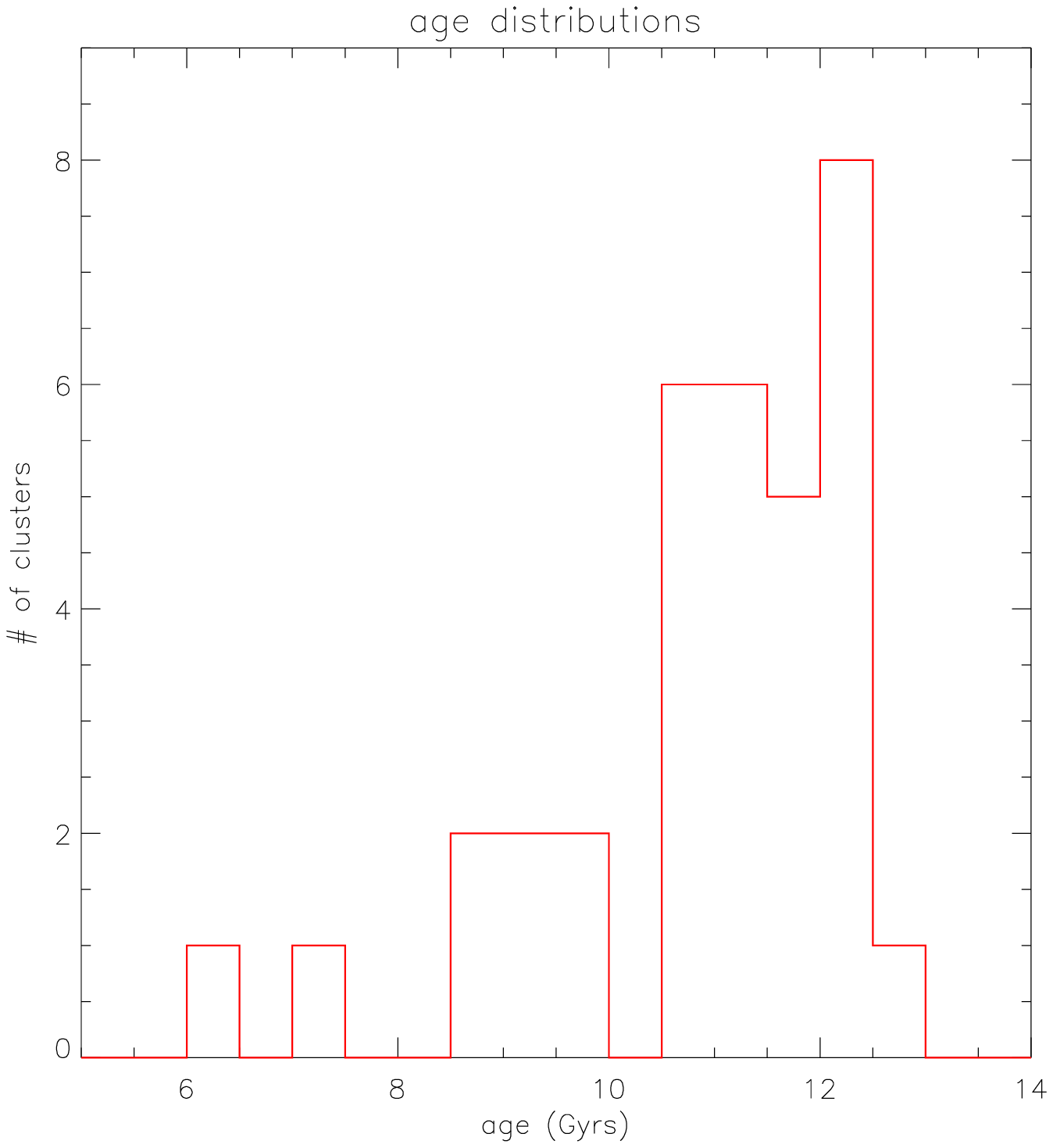}
\includegraphics[width=2.75in,angle=0]{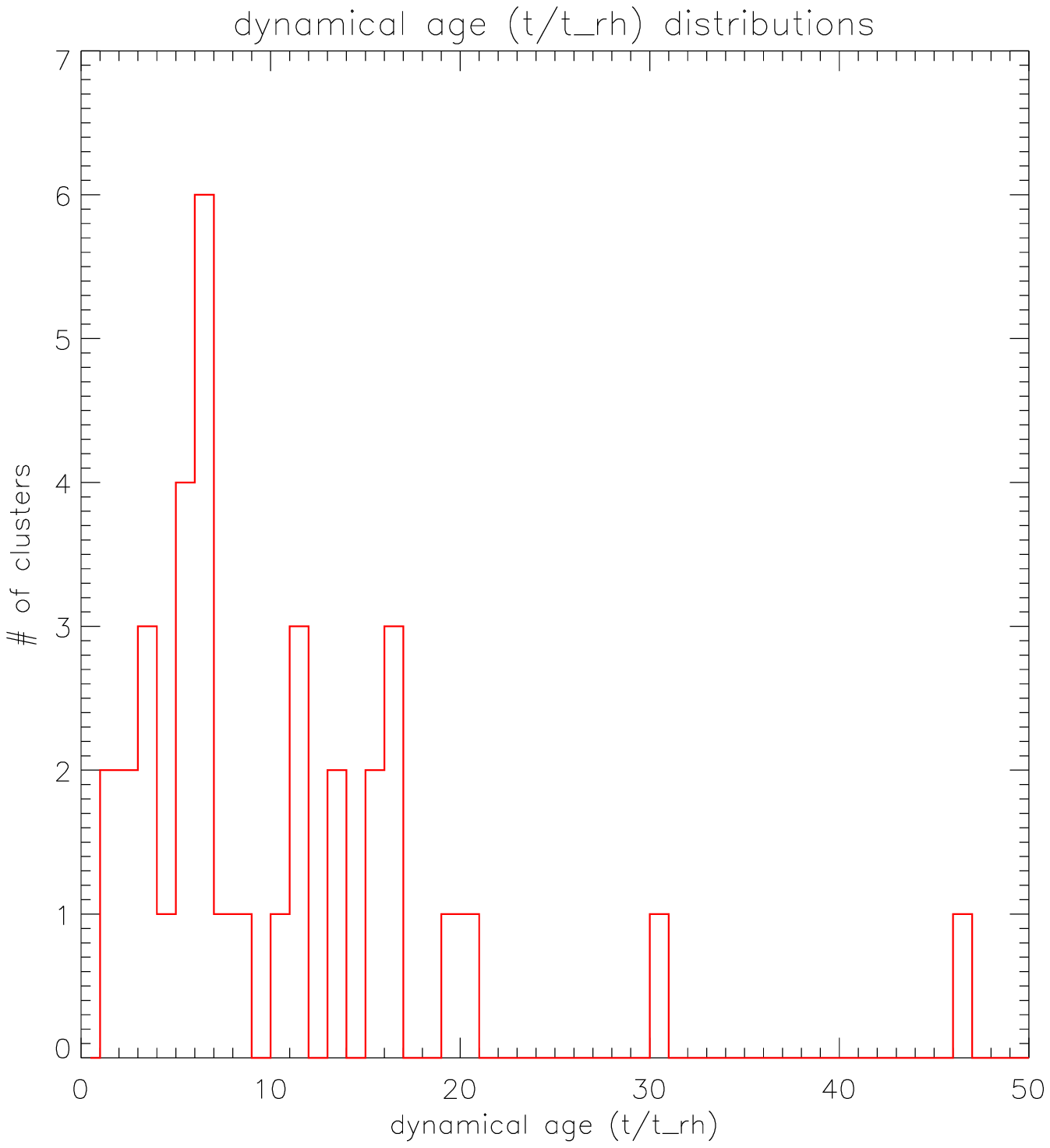}
\includegraphics[width=2.75in,angle=0]{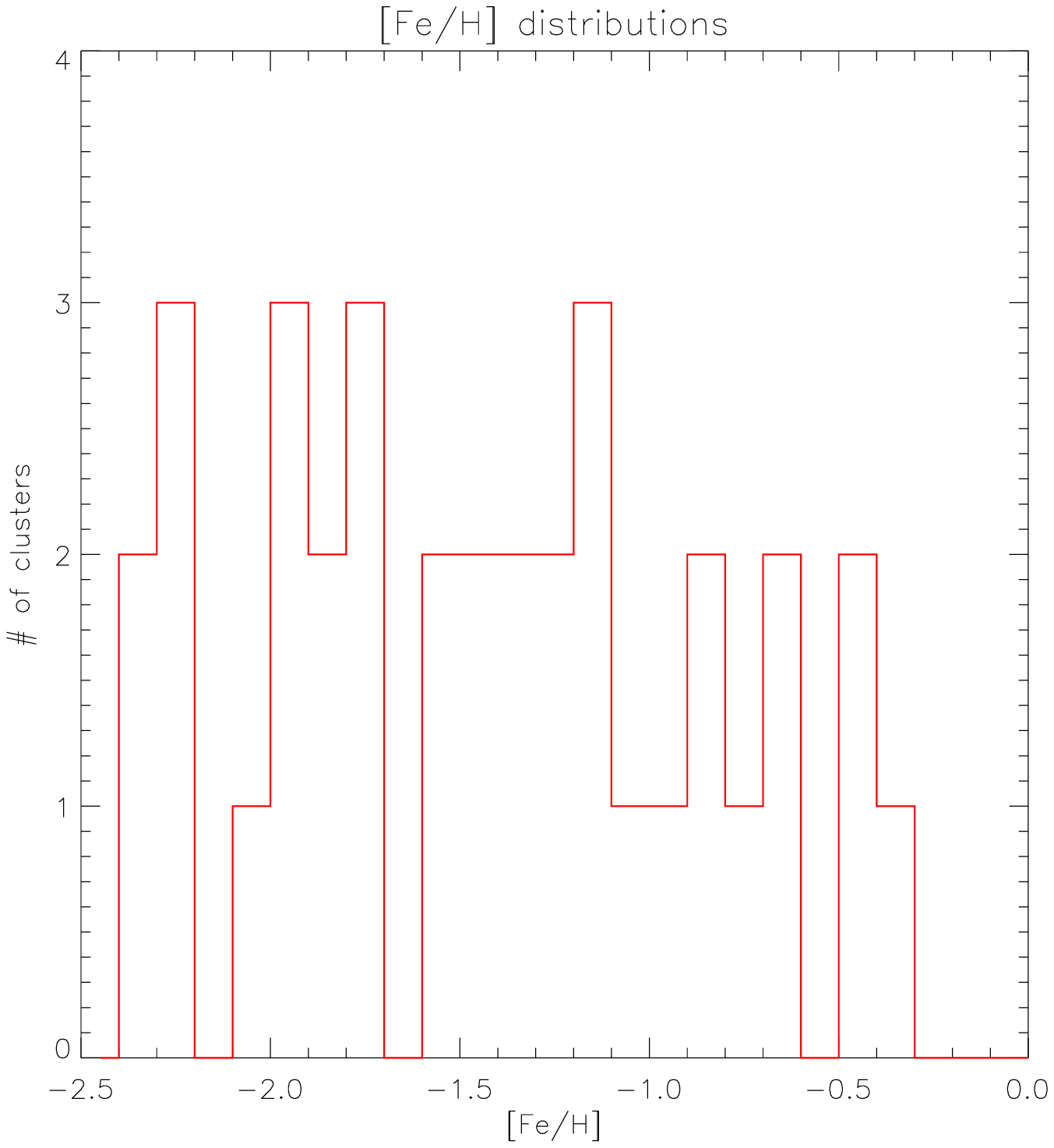}
\includegraphics[width=2.75in,angle=0]{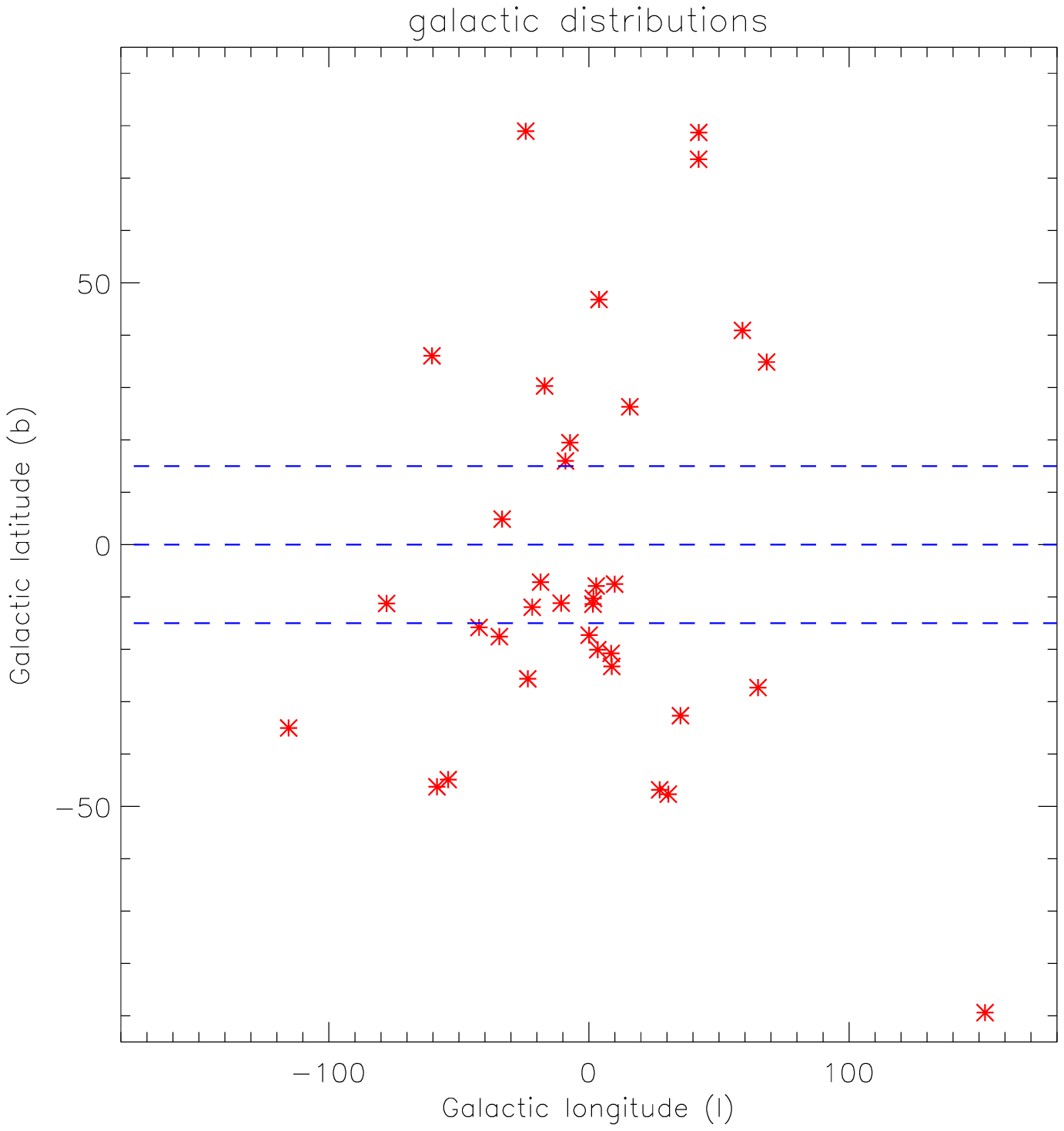}
\caption[Distributions of properties for the sample]{Distributions of properties for each globular clusters in the sample. Upper left: ages distribution. Upper right: dynamical ages distribution.
 Lower left: [Fe/H] distribution. Lower right: galactic position distribution.}
\label{fig:basic}
\end{figure}

Table ~\ref{table:basic} shows the basic information for each cluster
in this sample, including Galactic longitude (l), Galactic latitude (b), metallicity ([Fe/H]),
foreground reddening E(B-V), absolute visual magnitude $M_V$, collapsed
core (y/n), core radius ($r_c$), half light radius ($r_h$), log relaxation time at half mass radius
($\log t_{rh}$), age ($t_{Gyr}$), and dynamical age ($t_{Gyr}/t_{rh}$). The dynamical ages in this sample range from dynamical young clusters ($t_{Gyr}/t_{rh} = 1.5 $)
to dynamical old one ($t_{Gyr}/t_{rh} = 46.4 $)(see upper right panel in Figure ~\ref{fig:basic}). The
metallicities  in this sample range from metal poor cluster ($[Fe/H] = -2.37$)
to metal rich one ($[Fe/H] = -0.32$)(see Figure ~\ref{fig:basic}, lower left panel).
Figure ~\ref{fig:basic} lower right panel shows the distribution of those clusters relative
to the Galactic disk plane, among which 9 clusters are within the $\pm 15^{\circ} $ Galactic latitudes,
and are potentially affected by non-member stars.

Table ~\ref{table:obs log} shows the HST ACS observation log for each cluster,
including filter type, number of exposures used, exposure time per frame, and
data set ID. We only used the long exposure frames here and did not include those
short exposure frames, because we are only interested in the CMDs
below the turn-off point.

\section{Data Reduction and Photometry}\label{sec:reduction}
We retrieved the HST ACS archival data on F606W and F814W bands for each globular cluster, employing the most recent calibration frames on the fly.
A drizzled imaged was produced ({\it Multidrizzle} was applied to all {\it FLT} images), with care taken to achieve accurate alignment of the images.
Standard practices were applied for bad pixel masking and exposure calibrations.

\begin{figure}[tp] \centering
\includegraphics[width=2.75in,angle=0]{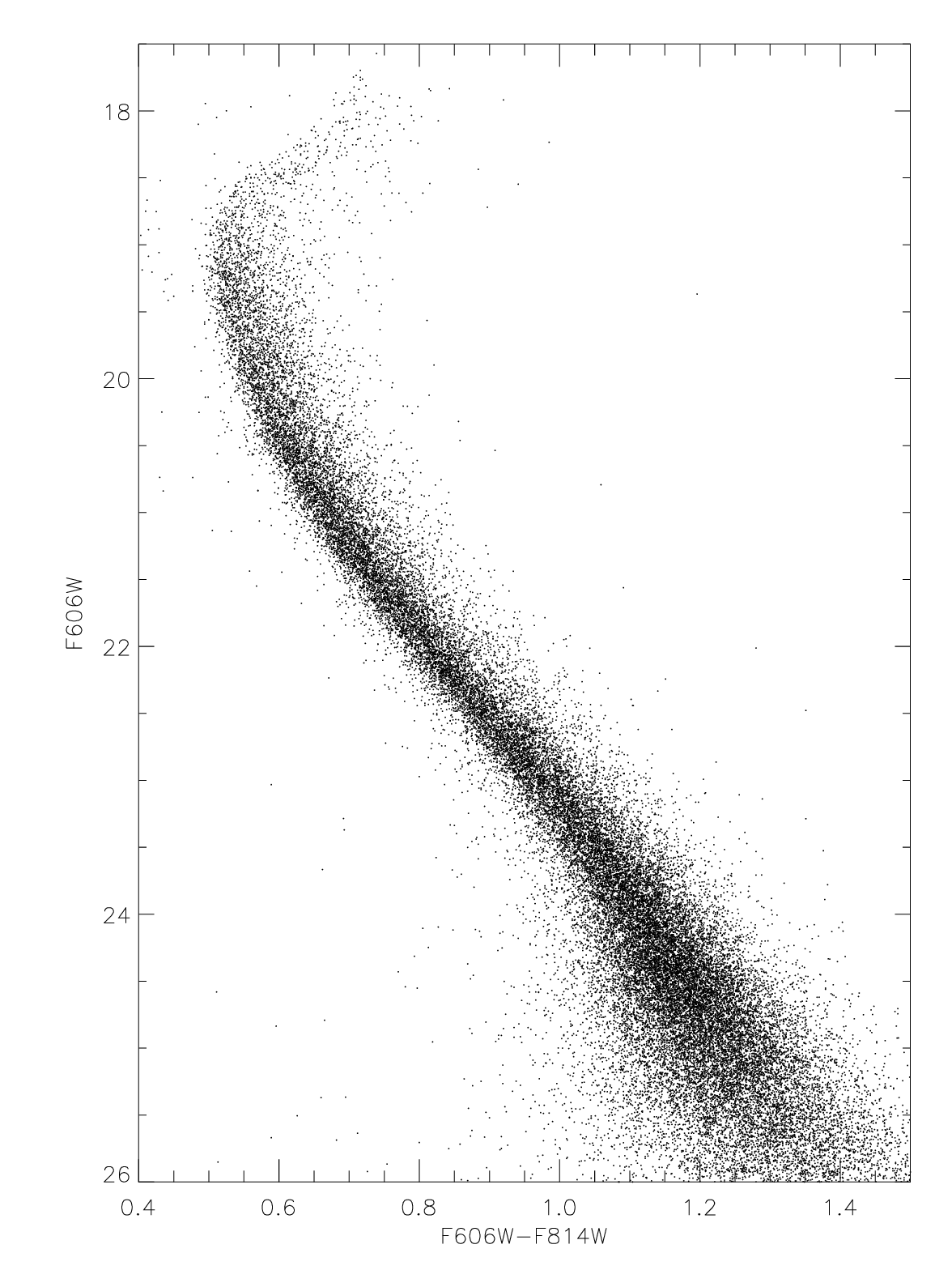}
\includegraphics[width=2.75in,angle=0]{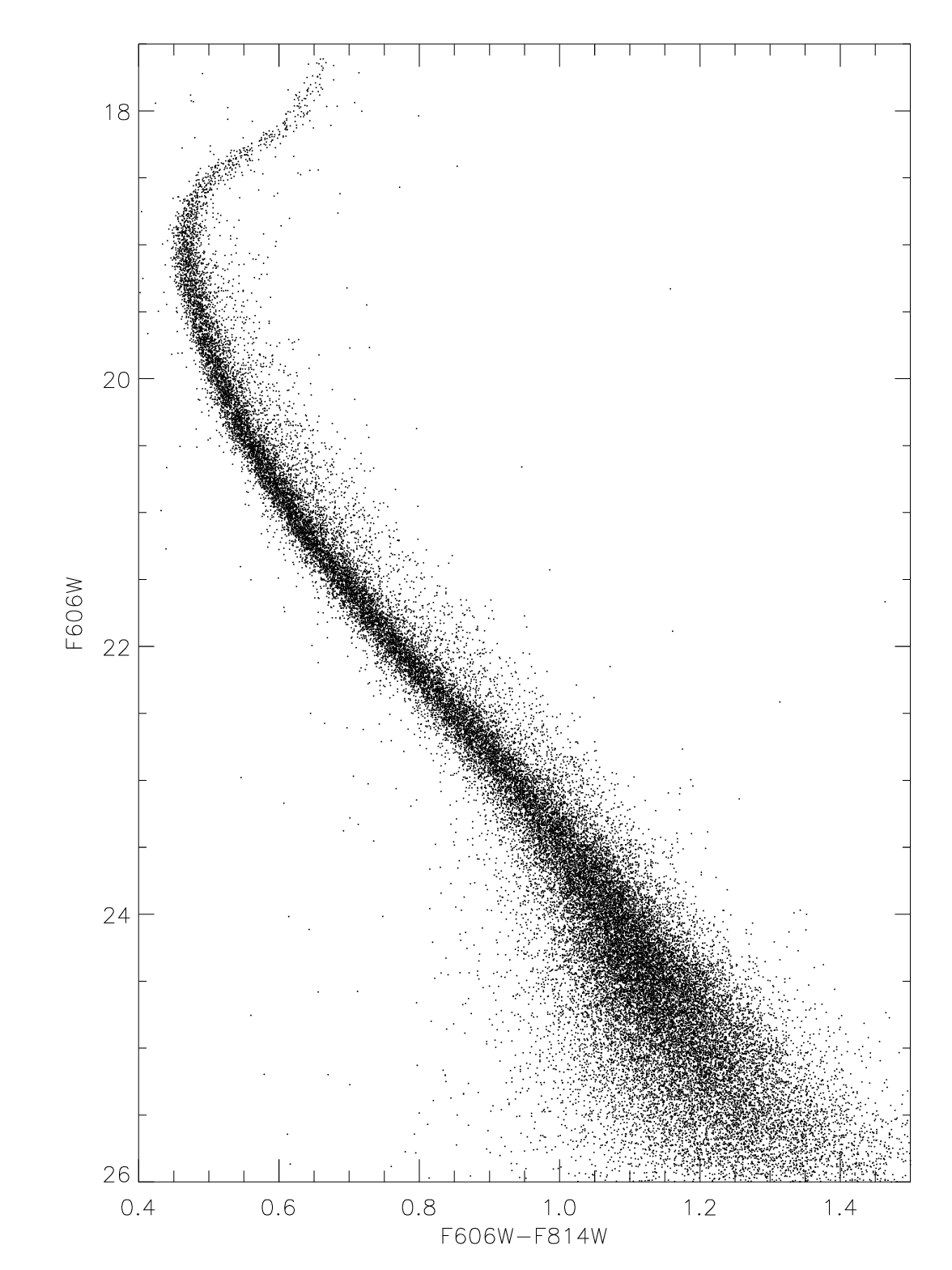}
\caption[Comparisons of the CMD quality between Dolphot V1.0 and V1.2]{Comparisons of the quality of the CMDs for NGC 4590 between different version of Dolphot, V1.0 (left), V1.2 (right). The latter version shows significant
improvement on the accuracy of the stars near the turn-off point, and shows a narrower spread
on the main-sequence.}
\label{fig:dolphot}
\end{figure}

The extraction of stellar magnitudes from the many point sources in these crowded fields depends on the algorithm used.  In these fields, especially the crowded central parts of clusters, a star often lies on the wings of another star and these wings are not azimuthally symmetric, which complicates the subtraction.  We used two versions of Dolphot \citep{dolphin00}, an automated CCD photometry package for general use and for HST data (ACS, WFPC2, and WPC3).
The later version uses a more refined library for the point spread function of point sources and we found significant improvement in the quality of PSF photometry for Dolphot V1.2 relative to V1.0.
This can be found in Figure ~\ref{fig:dolphot}, where the CMD for NGC 4590 near the turn-off point is well-defined and the main sequence is narrower by using Dolphot V1.2 (right panel) compared to Dolphot V1.0 (left panel).
The more accurate photometry leads to reduced uncertainties in the binary fraction determinations and allows us to probe to smaller binary mass ratios.
From the output photometry files, the error weighted average magnitudes for both the F814W and the F606W filters were used, with the zero point set to the VEGA system.
More details about the data reduction process can be found in \citet{ji2011}.
We note that in the work of \citet{milone12}, a different extraction routine was developed for their program which appears to lead to slightly smaller errors.

\begin{figure}
\includegraphics[width=6in,angle=0 ]{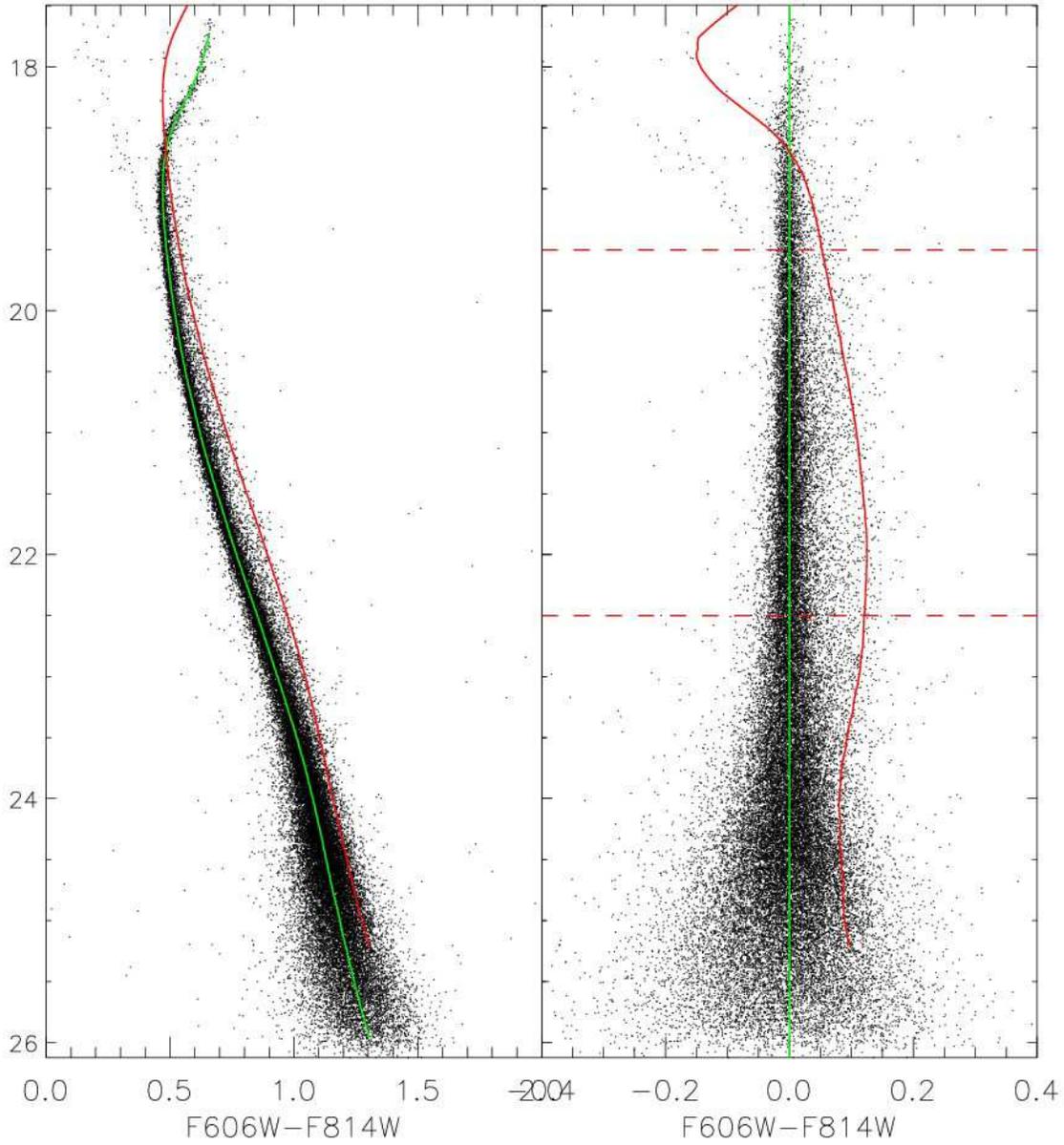}
\caption[Example for the observed CMD and the straightened CMD: NGC 4590]{An example of the observed CMD (left) and the straightened CMD (right) for NGC 4590. The solid green line is the defined MSRL. The solid red line is the equal-mass binary population. The binary fraction is estimated within the range of the dashed red lines.}
\label{fig:cmd4590}
\end{figure}

Not all parts of the resulting color-magnitude diagram are equally useful and only a limited range is used.  We chose stars one magnitude below the turn-off point, since close to the turn-off point, binary sequences turn to merge with the main sequence and lose all of the binary information.
At increasingly fainter magnitudes, the photometric uncertainty becomes large enough that those stars are not useful in constraining binary fraction models.
Typically, this occurs about four magnitudes below the turn-off point.
An example of these effects is shown in the CMD of NGC 4590 (Figure ~\ref{fig:cmd4590}). The red dashed lines indicate the portion of CMD we used during the analysis (other lines are discussed below).

\section{Main Sequence Definition and Artificial Star Tests}\label{sec:artificial}
Without knowing the photometric errors, completeness, and the rate of superposition of stars, we cannot have accurate measurement of binary fractions from the CMD. The
artificial star tests, however, can help us understand the photometric errors and accuracy, as well as completeness (or star recovery percentage) at different region of clusters and at different magnitudes. It can also simulate the rate of superposition of stars if performed properly.

\subsection{Defining the Main-sequence Ridge Line}
To perform artificial star tests and later CMD analysis, we need to define the main-sequence ridge line (MSRL) of the observed CMD first.
We constructed the MSRL from the data instead of fitting theoretical isochrones to the CMD, as the latter method is not sufficiently accurate to define the ridge line. To define the MSRL from the CMD, we used a moving box method. In each step, we defined a box with a height of 0.1 magnitude in the F606W axis, and a width of the entire range in the color (F606W-F814W) axis. Then we searched the peak value of the color histograms in that box with a color bin-size of 0.0015 magnitude. The color value at the peak and the middle point of the box in the F606W axis direction are the defined MSRL for stars in that box.
The process was repeated in increments along the Main Sequence and the resulting
line was then smoothed with the moving average method with a period of 50.
In Figure ~\ref{fig:cmd4590}, the green line is the defined MSRL.

\subsection{Performing the Artificial Star Tests}
\begin{figure}
\includegraphics[width=3in,angle=0]{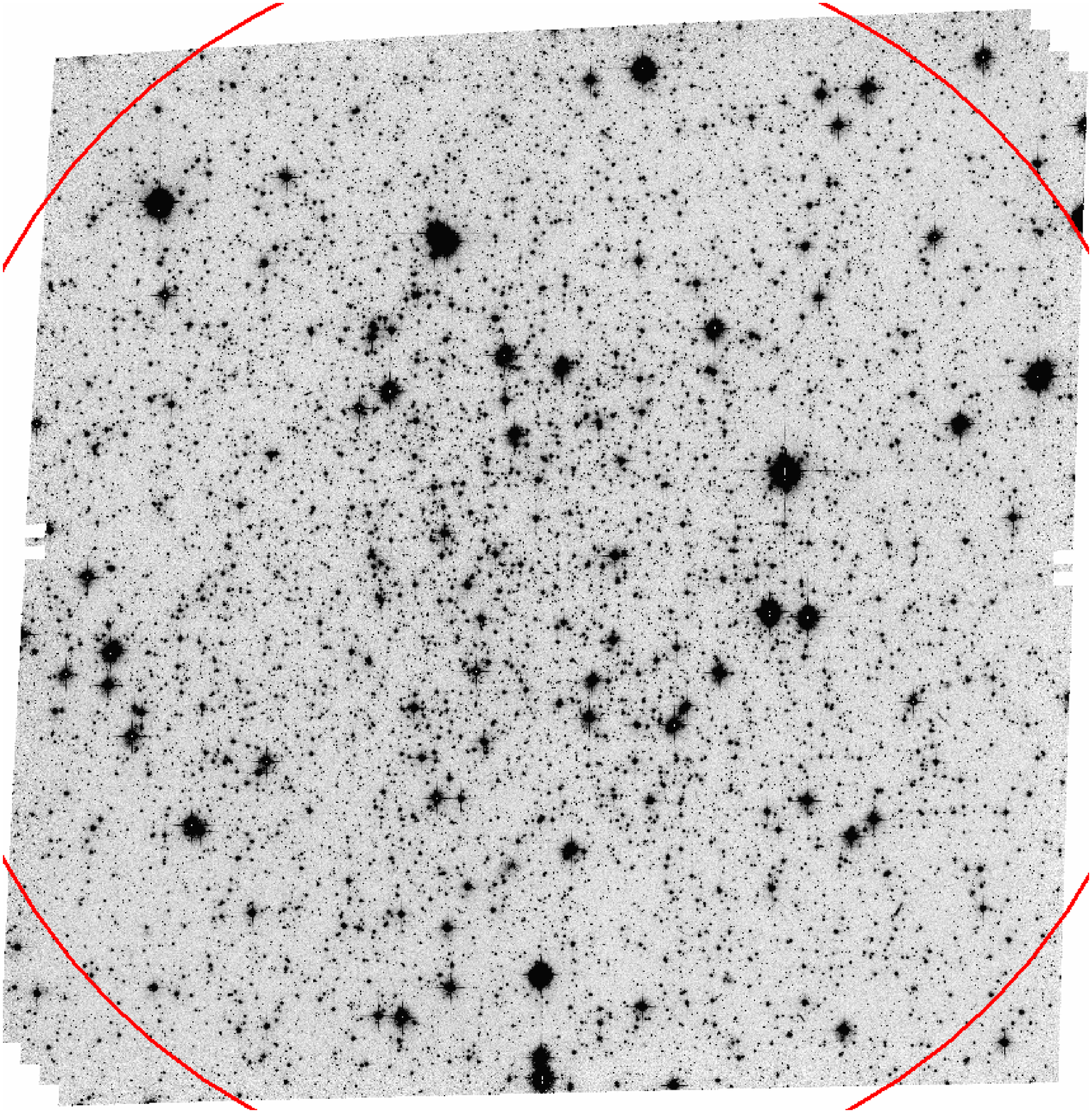}
\includegraphics[width=3in,angle=0]{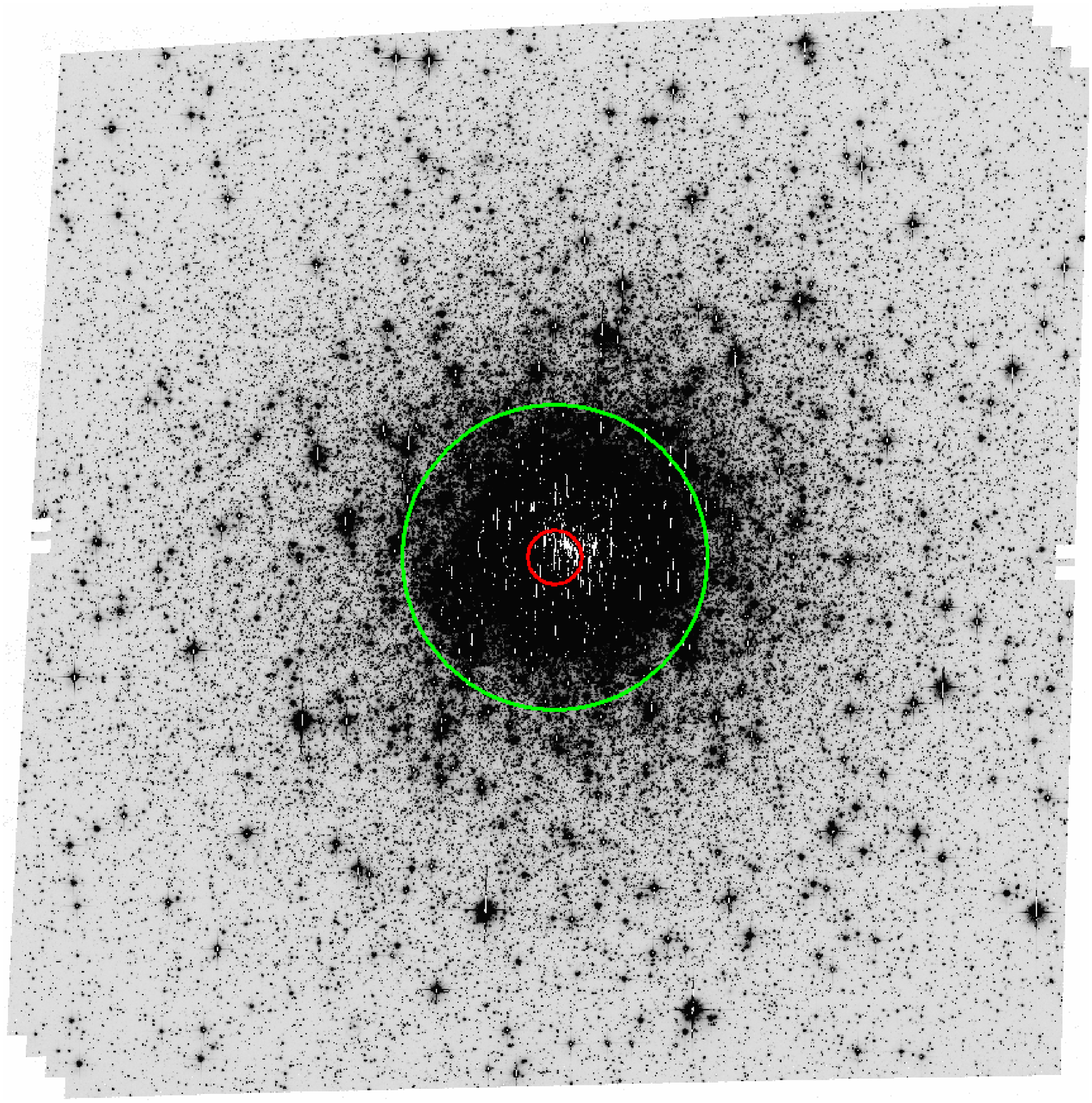}
\includegraphics[width=3in,angle=0]{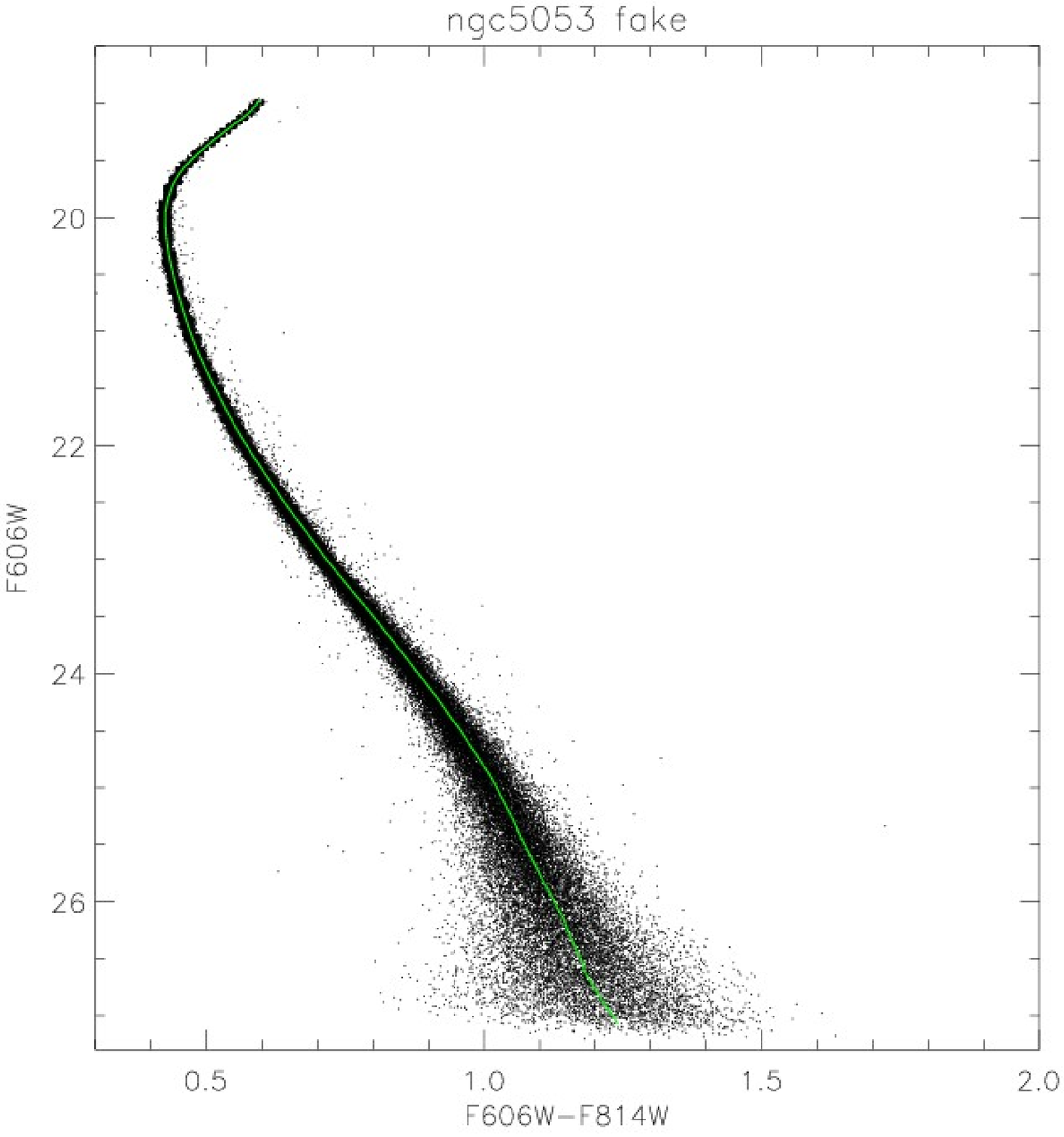}
\includegraphics[width=3in,angle=0]{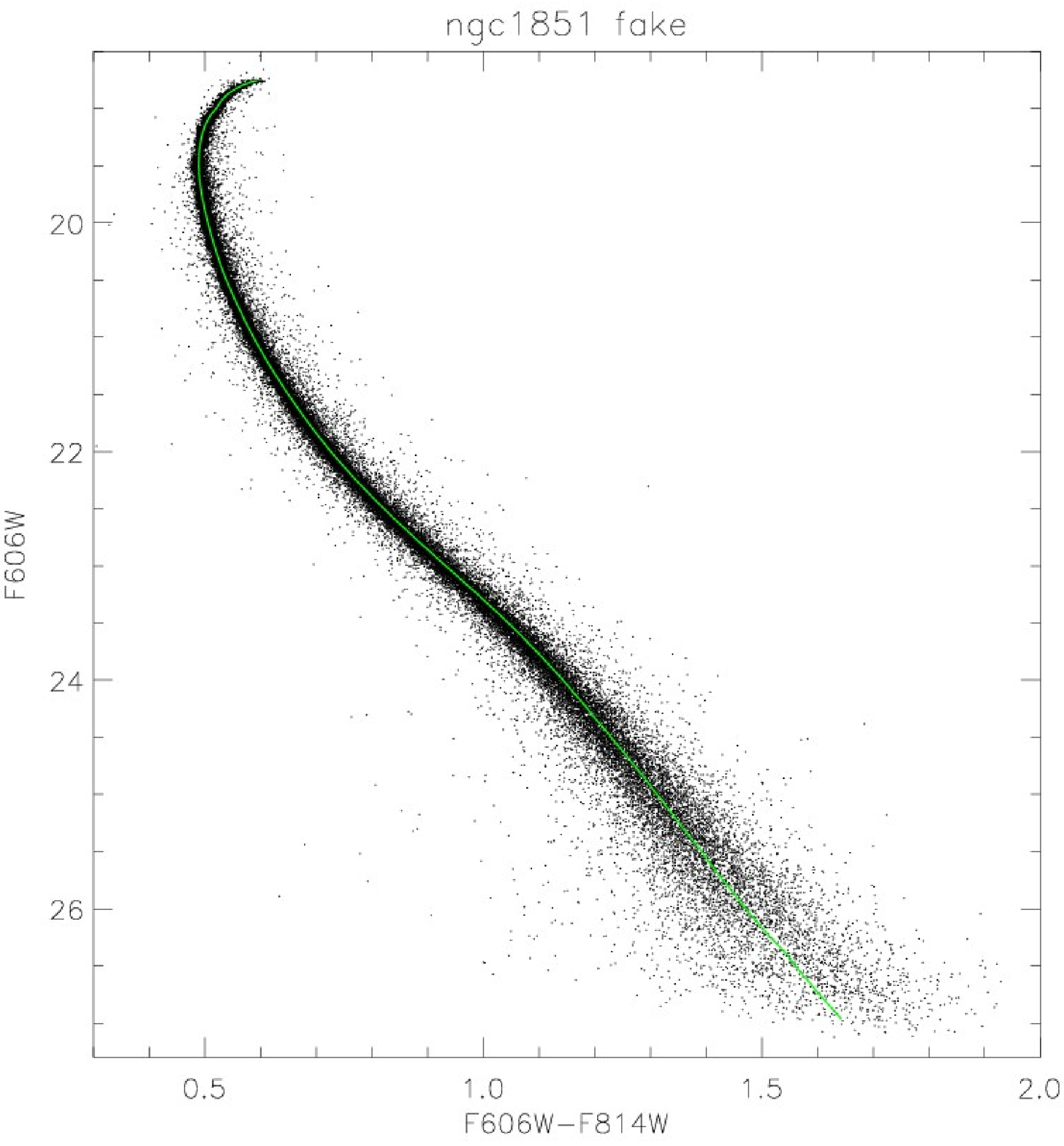}
\caption[Examples of images and fake CMDs: NGC 5053 and NGC 1851]{Two examples of the artificial star tests. Left column: low stellar density cluster NGC 5053. Right column: high stellar density cluster NGC 1851. Upper panel: the observed ACS images for these two clusters. The red circle represents core region while the green one is for the half light region. Lower panel: the input fake stars (green lines) and the recovered fake stars (black dots) on CMDs.}
\label{fig:arti}
\end{figure}

With the defined MSRL, we performed artificial star tests in Dolphot.
The F606W magnitudes of those fake stars were randomly generated while the F814W magnitudes were derived from the MSRL, so that all those fake stars are on the MSRL. The (X, Y) coordinates of those fake stars were randomly chosen from the area of the drizzled reference frame, but were not at the empty non-data CCD area. Each fake star in this list was added to each {\it FLT} frame one at a time, and was recovered using the same photometry process by Dolphot. The final photometric output file was screened using the same criteria as used to obtain the CMD, as described above; we added 100,000 fake stars for each image. Figure ~\ref{fig:arti} shows two examples of the artificial star tests, NGC 5053 (low stellar density cluster) and NGC 1851 (high stellar density cluster). The green line is the input fake stars, which are all on the MSRL. The scattered black dots are the recovered fake stars (69,995 stars for NGC 5053 and 52,205 stars for NGC 1851).

\subsection{Photometric Accuracy, Uncertainty, and Completeness}
From the input and recovered fake star lists, we studed the photometric accuracy and uncertainty.
We compared the photometry for two clusters, NGC 5053 (low stellar density cluster) and NGC 1851 (high stellar density cluster), as their HST ACS observations are very similar in exposure times: NGC 5053 (F606W: 340s$\times$5, F814W: 350s$\times$5) and NGC 1851 (F606W: 350s$\times$5, F814W: 350s$\times$5).

\begin{figure}[htbp] \centering
\includegraphics[width=2.5in,angle=0]{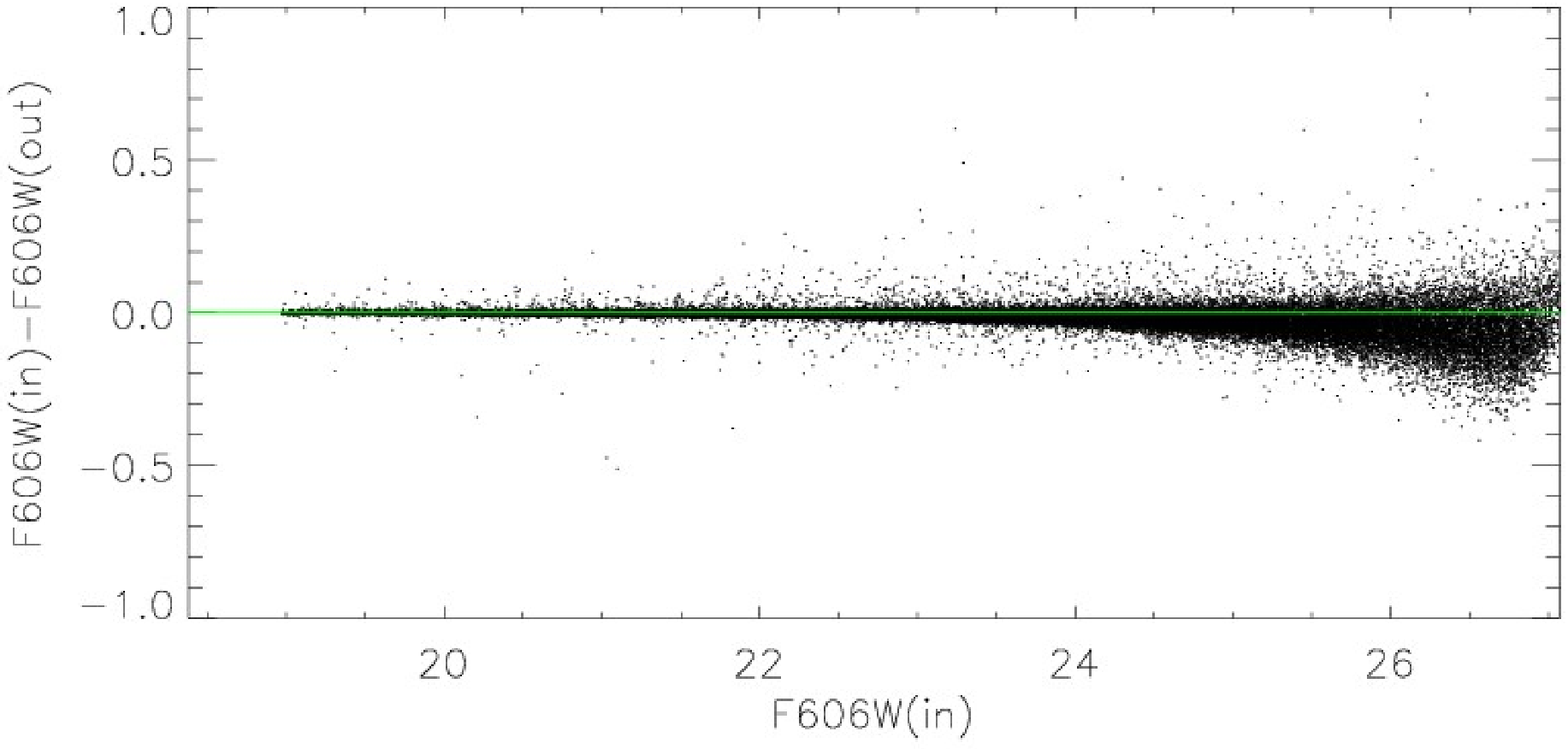}
\includegraphics[width=2.5in,angle=0]{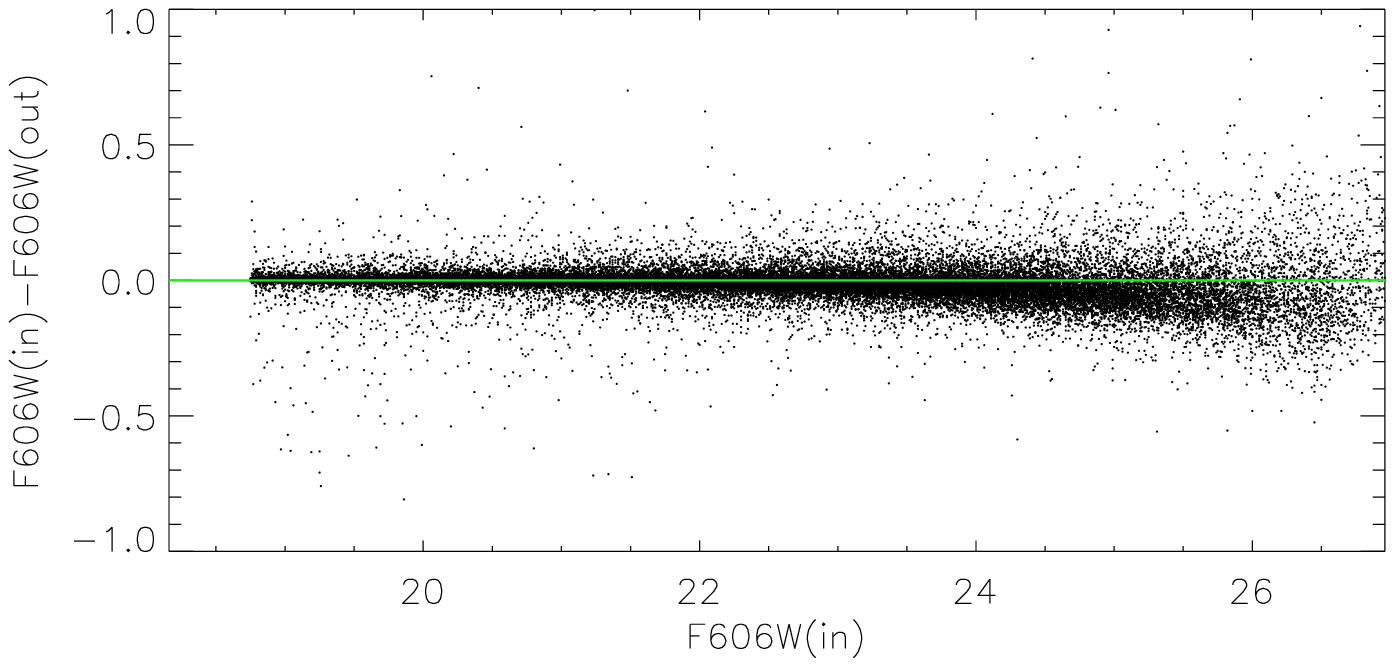}
\includegraphics[width=2.5in,angle=0]{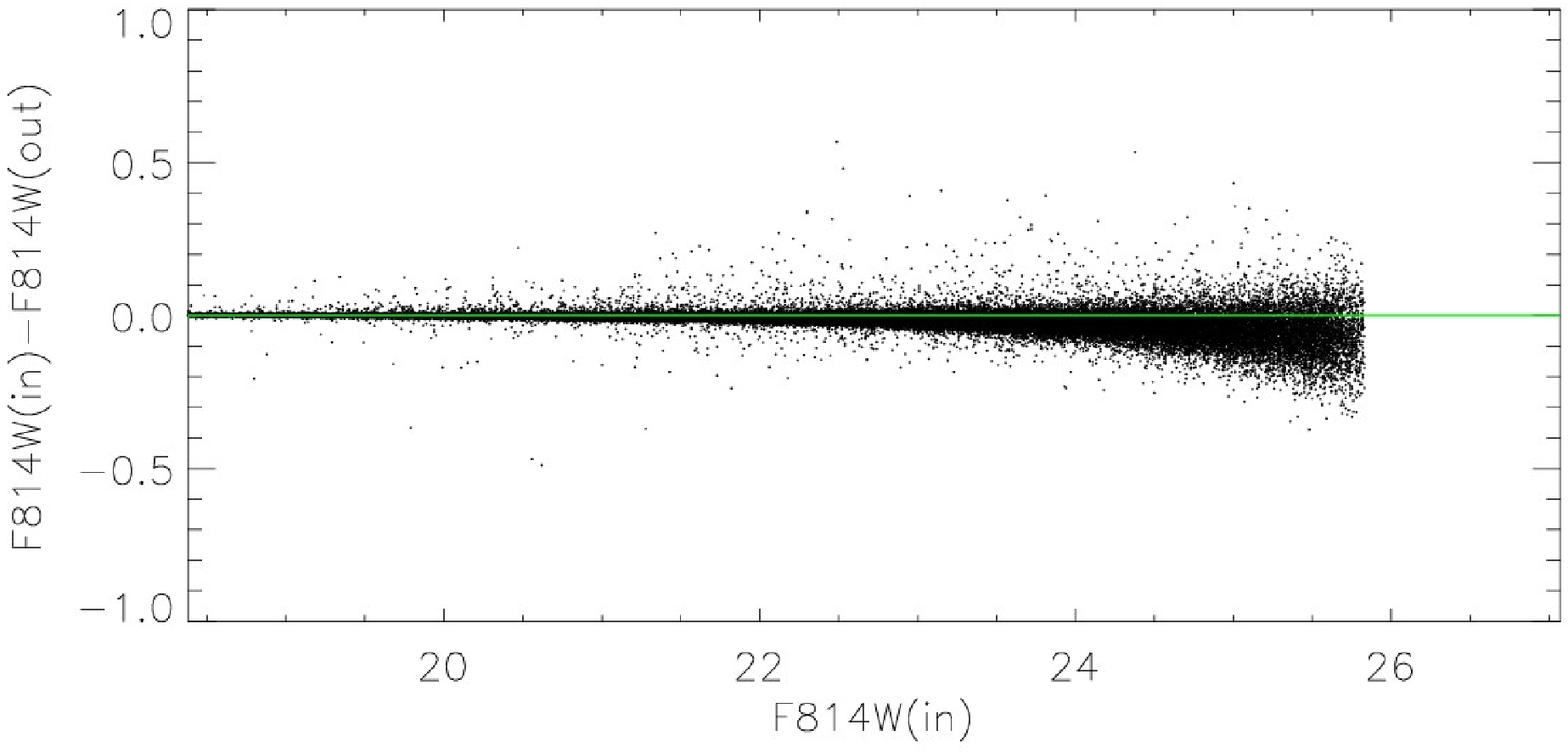}
\includegraphics[width=2.5in,angle=0]{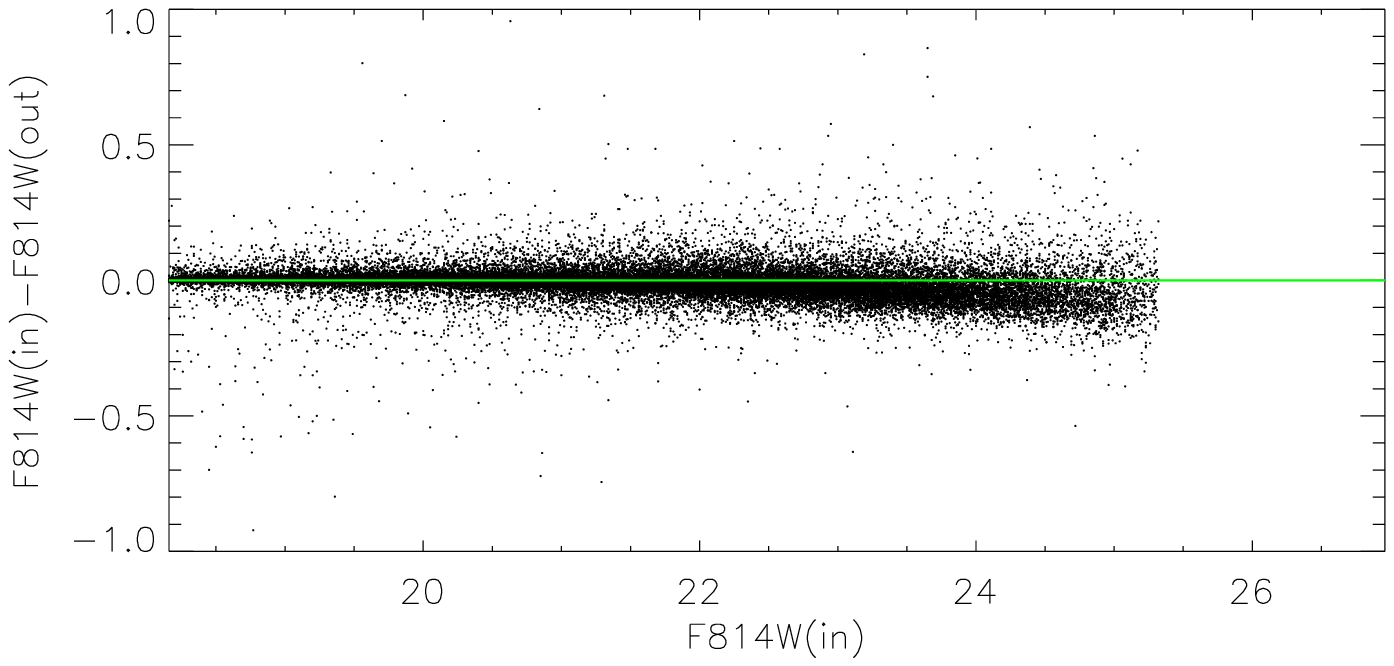}
\includegraphics[width=2.5in,angle=0]{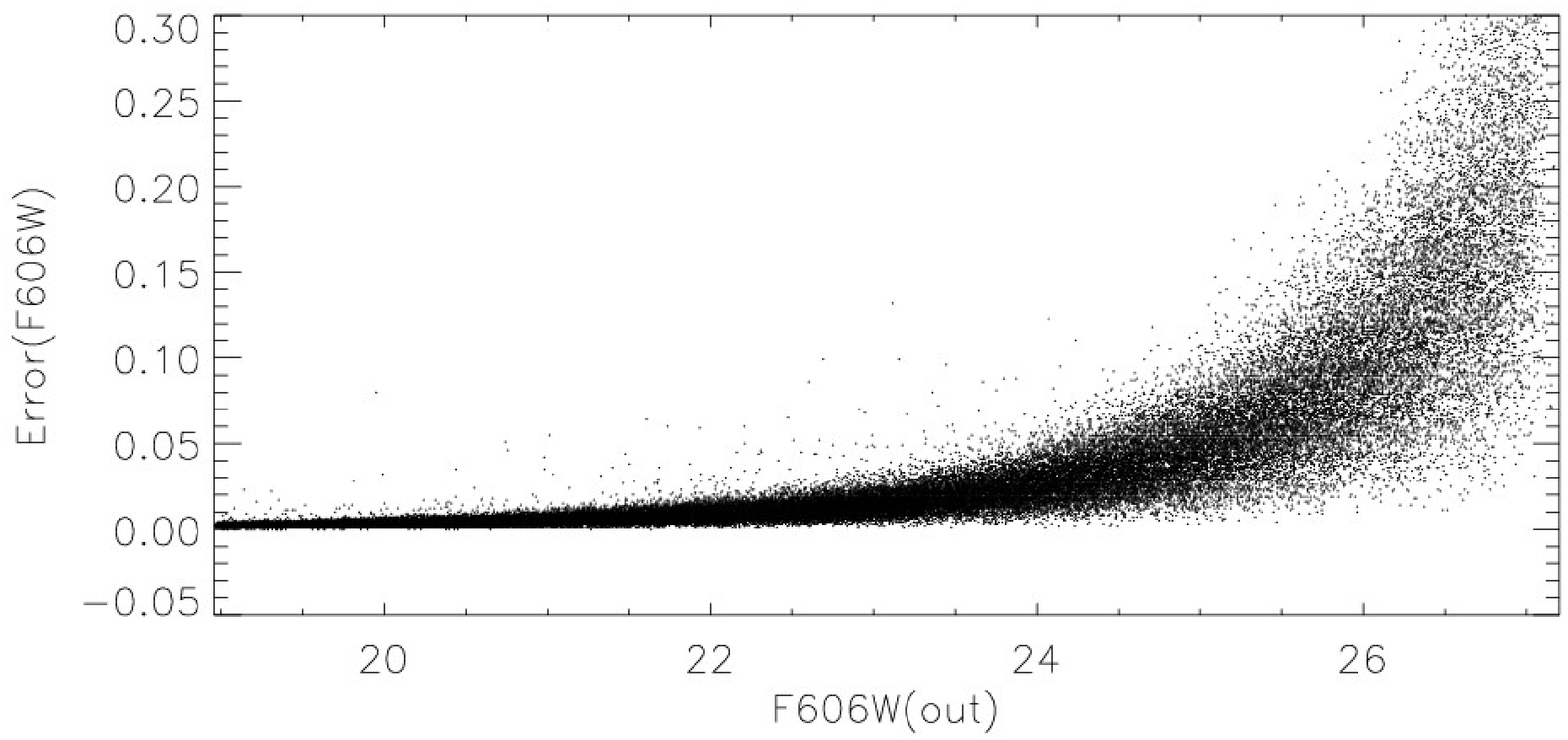}
\includegraphics[width=2.5in,angle=0]{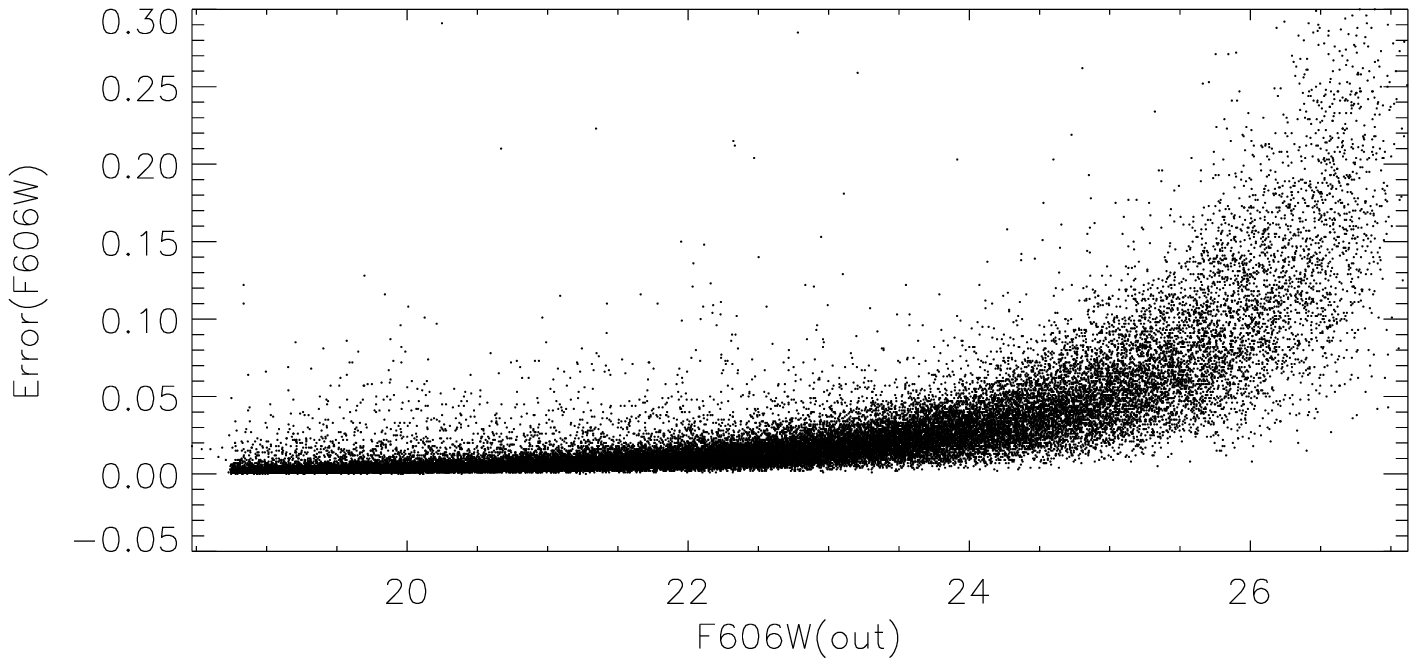}
\includegraphics[width=2.5in,angle=0]{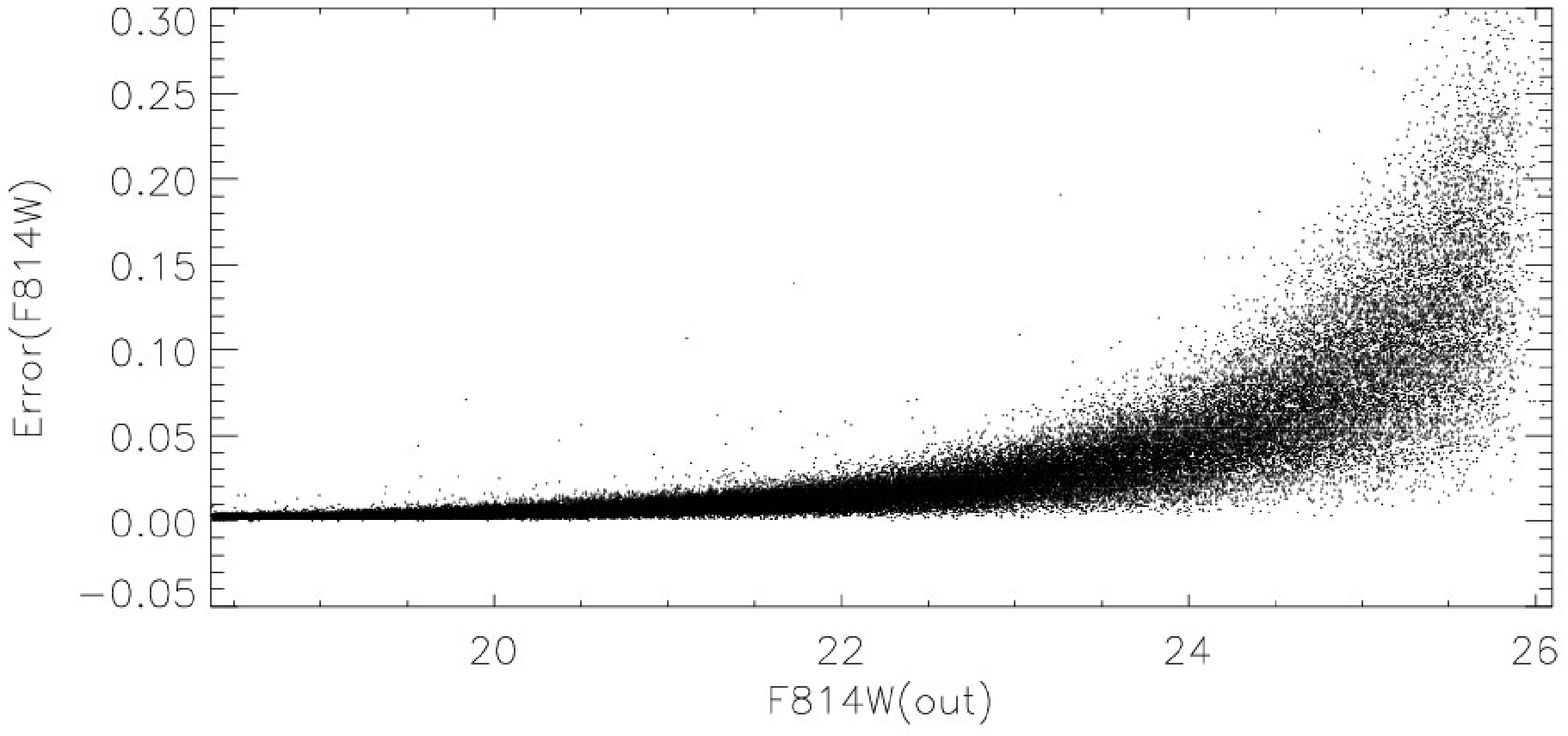}
\includegraphics[width=2.5in,angle=0]{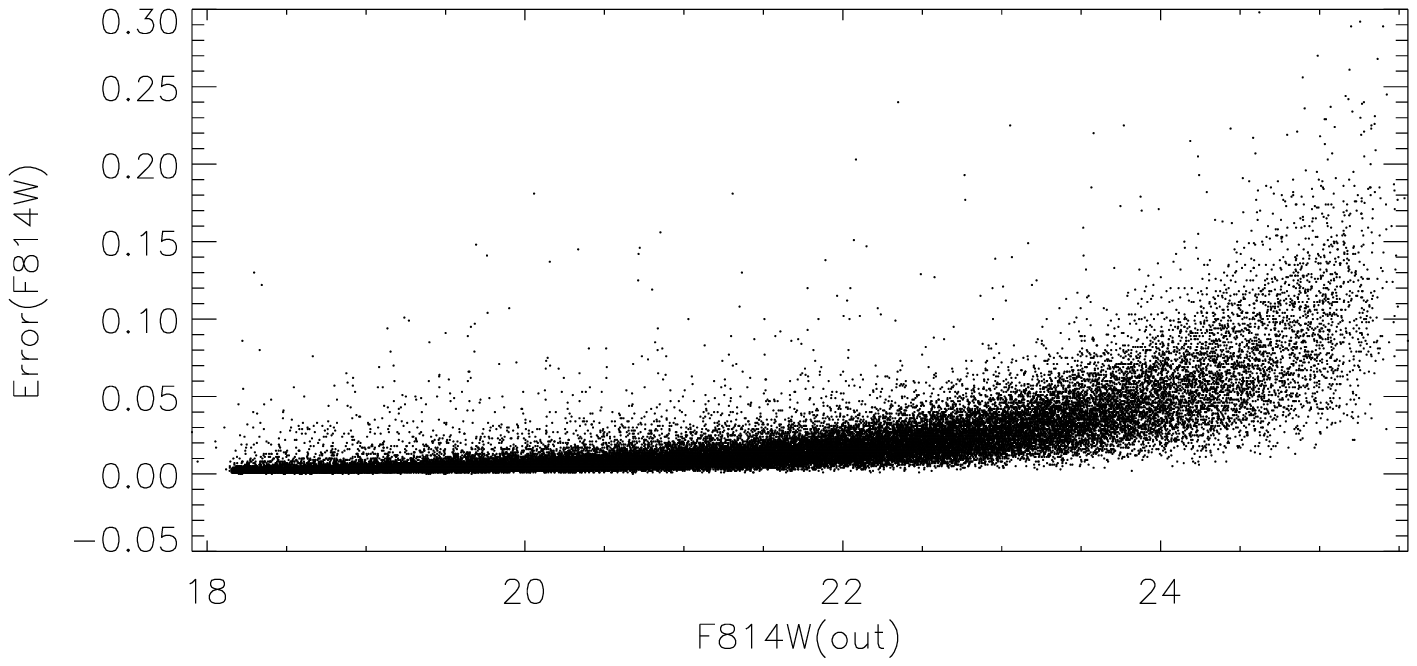}
\includegraphics[width=2.5in,angle=0]{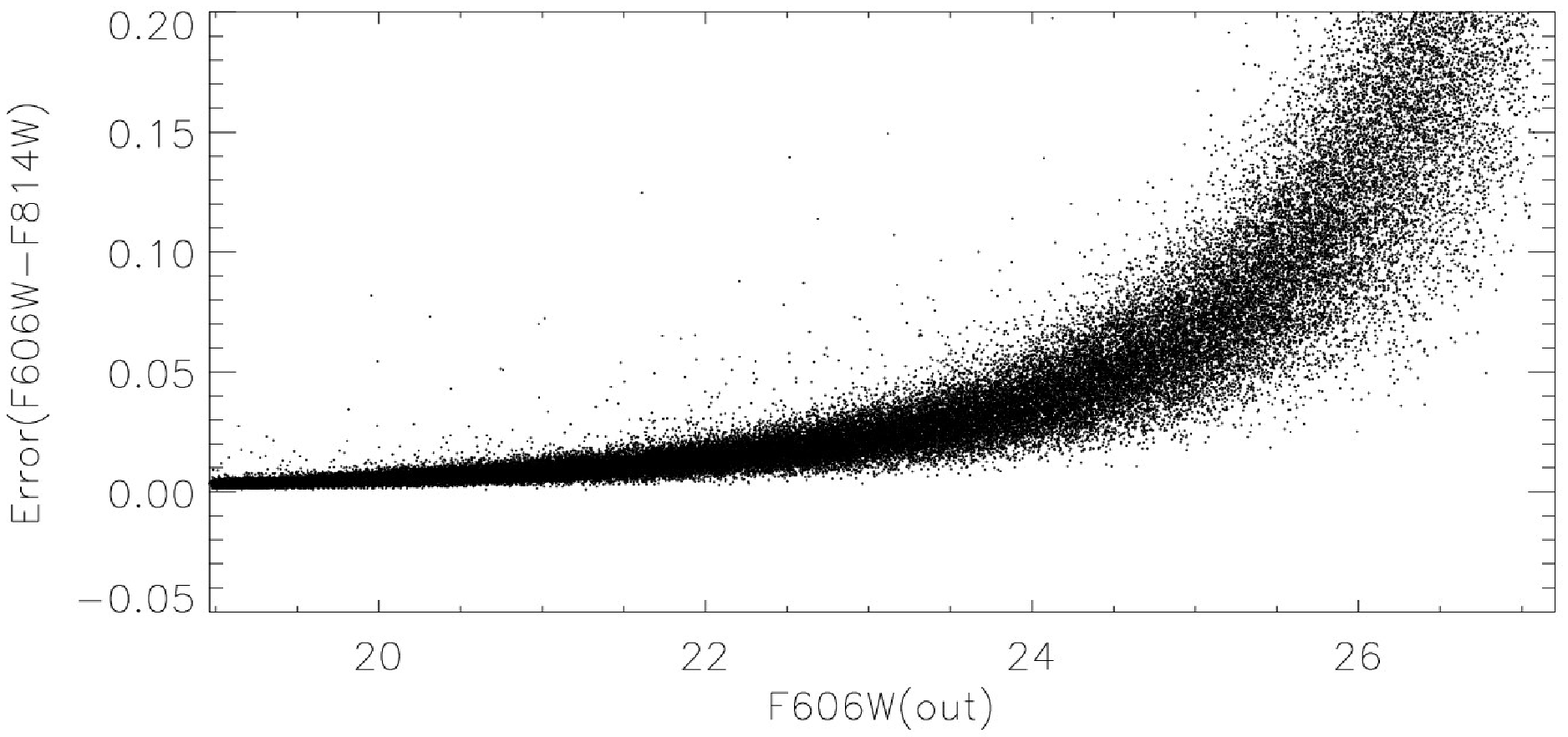}
\includegraphics[width=2.5in,angle=0]{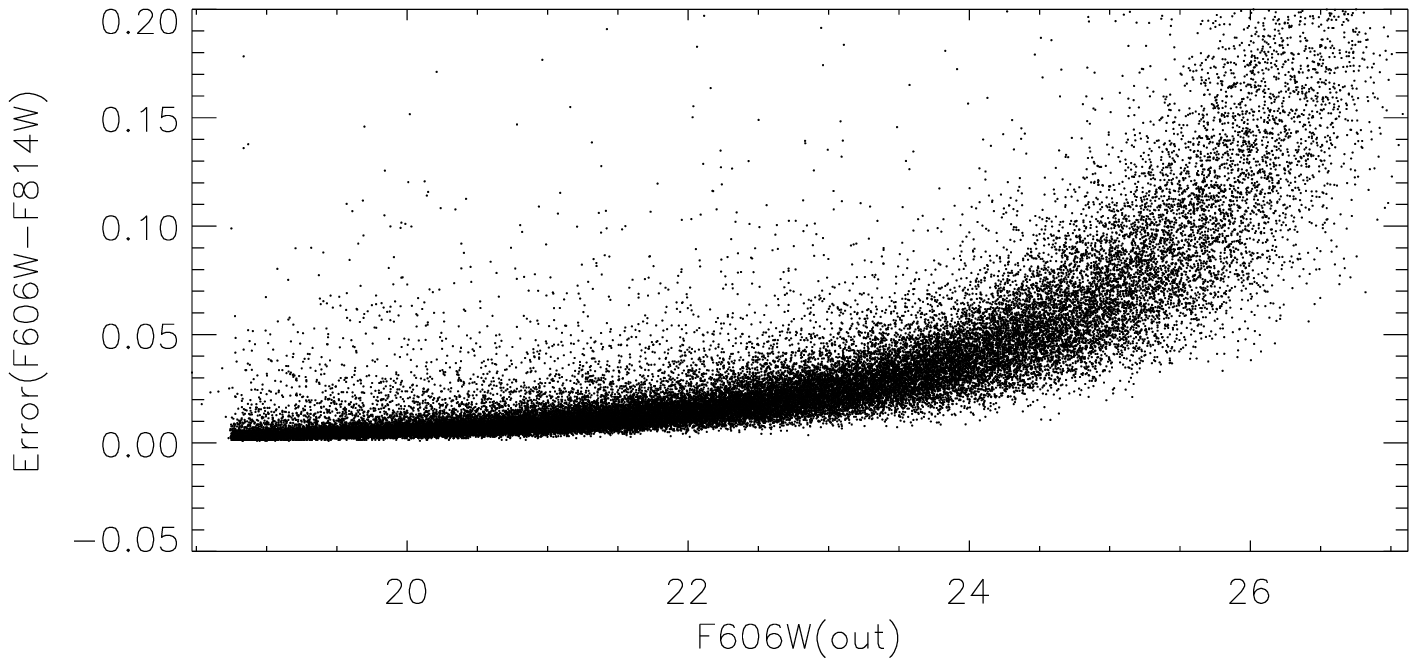}
\includegraphics[width=2.5in,angle=0]{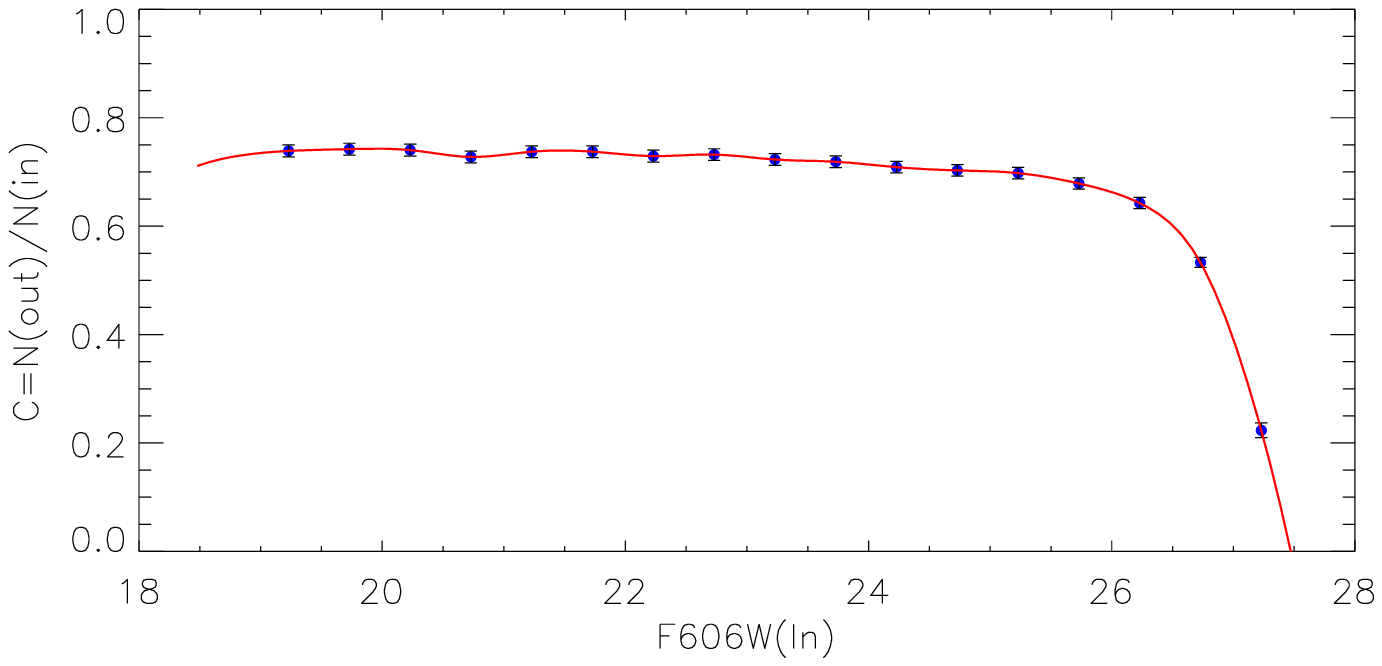}
\includegraphics[width=2.5in,angle=0]{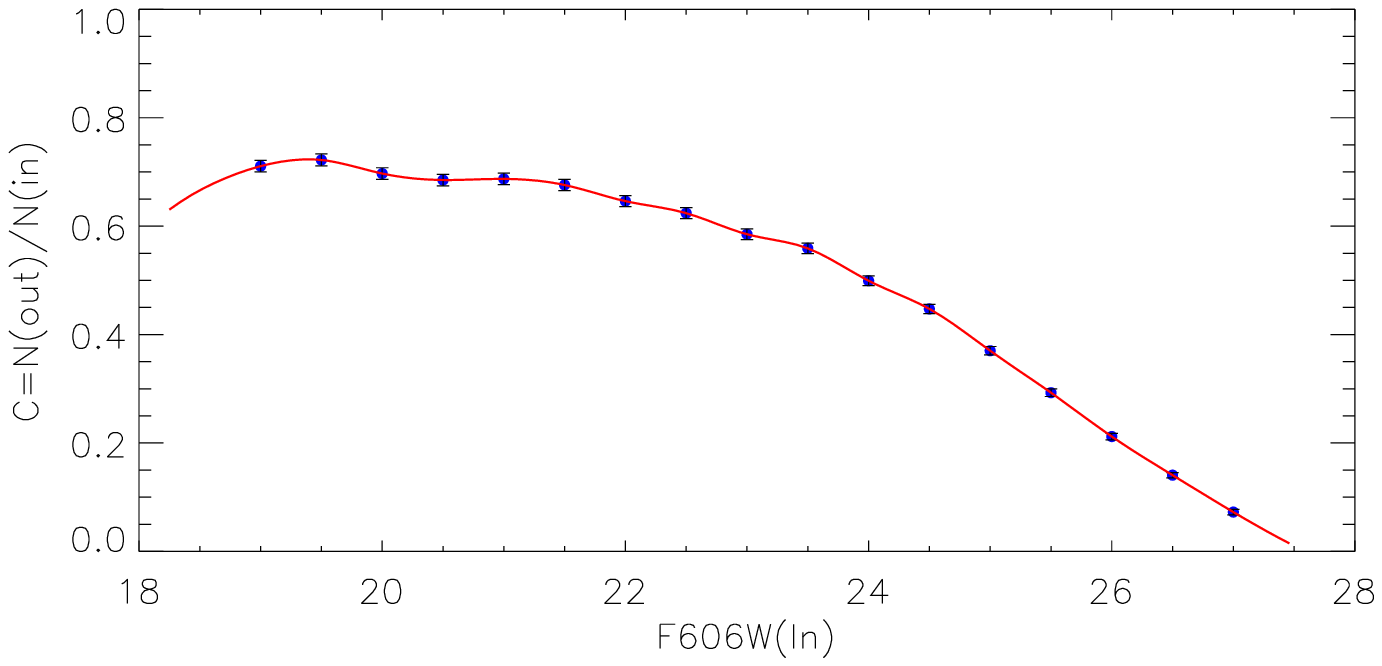}
\caption[Examples for the artificial star tests: NGC 5053 and NGC 1851]{Examples of the results for the artificial star tests: NGC 5053 (left column) and NGC 1851 (right column). Row 1 \& 2: differences of the input and recovered magnitudes for filter F606W and F814W. Row 3 \& 4: magnitude uncertainties for filter F606W and F814W. Row 5: color uncertainties. Row 6: completeness curve. }
\label{fig:err}
\end{figure}

In Figure ~\ref{fig:err}, we plot the magnitude differences for input and recovered stars for F606W filter (1st row) and F814W filter (2nd row). These plots indicate photometric discrepancies, and all of them showing differences within $\pm 0.2$ magnitudes. At fainter magnitudes, the magnitudes of the recovered stars tend to be fainter than the input ones. Row 3 and 4 in Figure ~\ref{fig:err} show the photometric uncertainties for F606W and F814W filters, respectively, while row 5 in Figure ~\ref{fig:err} shows the uncertainties in color (F606W-F814W).
From those plots we can see that most stars in NGC 1851 have similar photometric accuracy and uncertainties
as in NGC 5053, except that in NGC 1851 there are more stars with large errors, even for bright stars.

Row 6 in Figure ~\ref{fig:err} shows the completeness curve (recovery percentage), which is the ratio of
the total number of recovered stars (N(out)) to the total number of input stars (N(In)) in a particular magnitude interval.
The completeness curve for NGC 5053 is quite flat down to the 26.5  magnitude, indicating an even star recovery percentage over a broad magnitude range. However, for NGC 1851, the star recovery percentage is not evenly distributed, with a relatively higher recovery percentage for brighter stars than for fainter ones.
This can be seen from the fake CMDs in Figure ~\ref{fig:arti}, where there are fewer faint stars in NGC 1851 than in NGC 5053, even though they have similar exposure times. This indicates that faint stars are difficult to recover in dense stellar region due to high and non-uniform background caused by the multitude of stars.

\section{The High Mass-ratio Binary Fraction}\label{sec:high-q fb}
\begin{figure}\centering
\includegraphics[width=2.5in,angle=0 ]{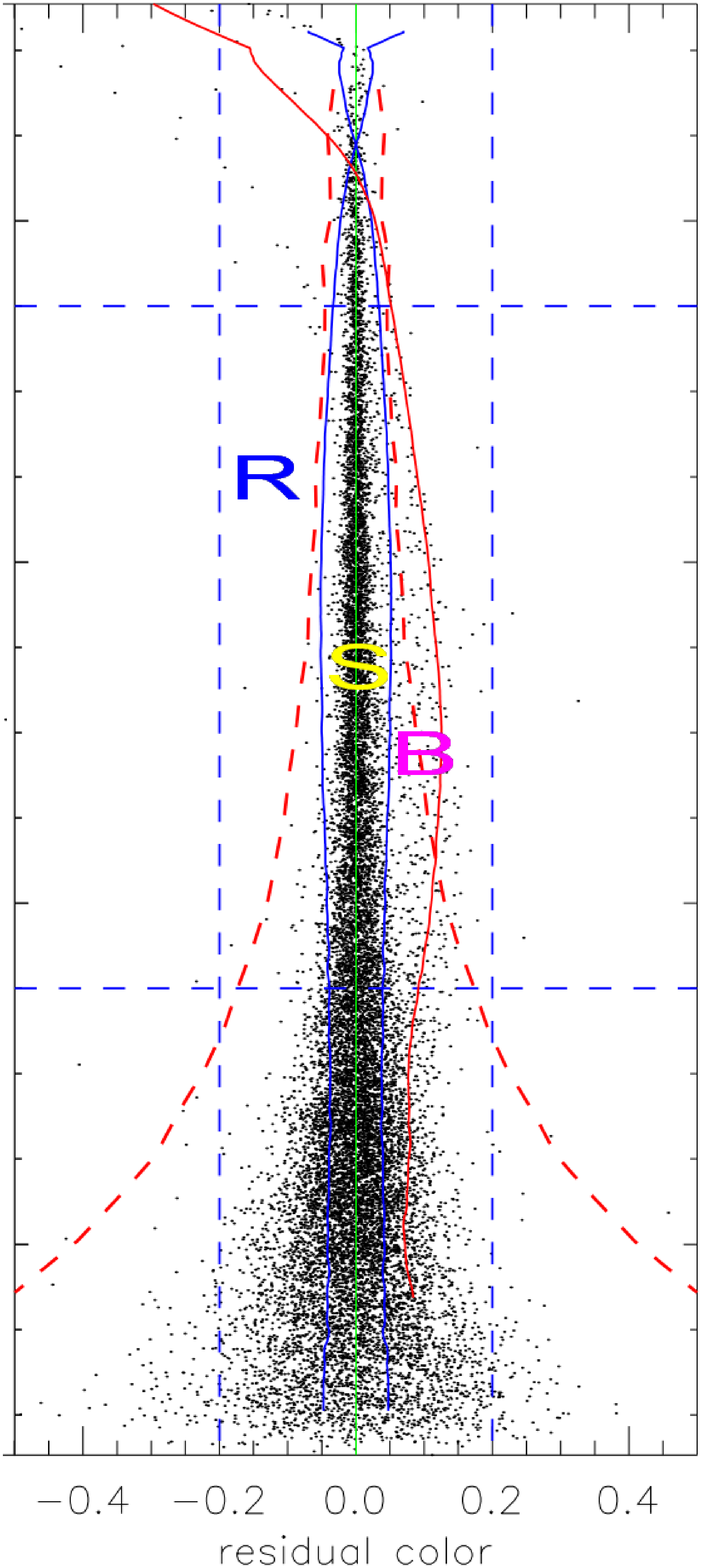}
\includegraphics[width=2.5in,angle=0 ]{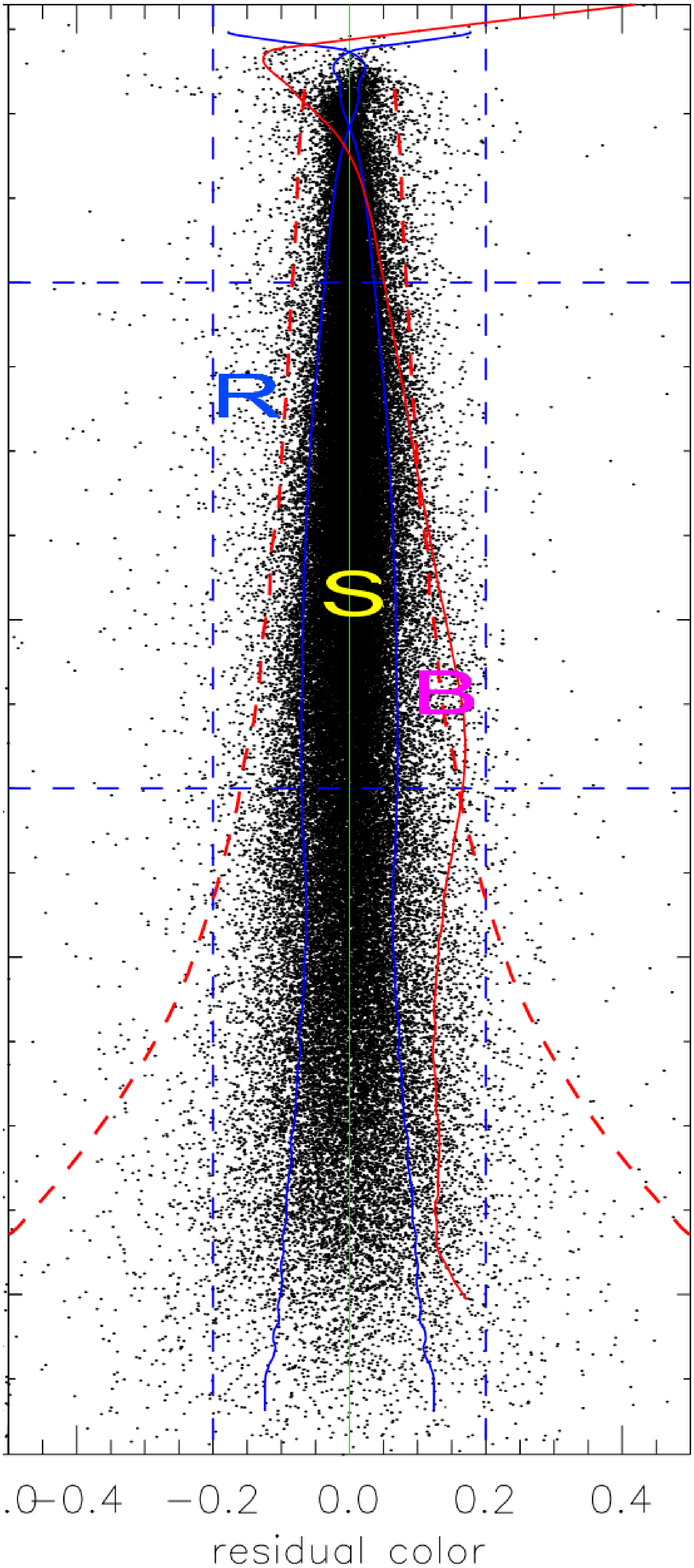}
\caption[Demonstration of the regions when measuring the minimum binary fraction for NGC 5053 and NGC 2808]{\footnotesize Demonstration of the regions when measuring the minimum binary fraction. The straightened color-magnitude diagrams are for NGC 5053 (left) and NGC 2808 (right). Left panel: the solid green line is the center of the main sequence. The solid red line is the equal-mass binary population. The solid blue curves are the binary population where the mass ratio equals 0.5 (right curve), and its symmetric one on the left. The dashed red lines are the $\pm 3\sigma$ photometric spread of color in comparison. The dashed blue lines are the upper and lower limits of usable stars for both F606W magnitude and residual color.
The main-sequence star (or single star) region $S$ is defined between the solid blue lines. The binary region $B$ is where the residual color is greater than the 0.5 mass-ratio binary line and less than 0.2. The residual region $R$ is where the residual colors are beyond the blue side of the left solid blue line but greater than -0.2. Right panel: same as left but for NGC 2808 with more broadened main sequence due to multiple populations in the main sequence (see \citet{Piotto07}). The 3 $\sigma$ photometric limit lines (dashed red lines) intersect with the equal-mass binary line, indicating that most binary signals are buried. }
\label{fig:region}
\end{figure}

A binary with a mass-ratio close to unity is easiest to identify because they binaries have the largest distances from the main-sequence ridge line. This is evident in an examination of the straightened CMD (i.e. the CMD minus the color of the MSRL), where the binaries are distributed to the right side of the main sequence below the turn-off point (Figure ~\ref{fig:region}). The maximum distance is from equal-mass binaries, and this distance varies with the shape of the MSRL (the green line in Figure ~\ref{fig:region}). The minimum separation from the main sequence is for binaries where the secondary is of low mass (mass-ratios approaching zero; see the blue solid line in Figure ~\ref{fig:region}). When the photometric errors increase, the main-sequence stars eventually spread to the equal-mass binary region, diluting the binary signature (see Figure ~\ref{fig:region}, right panel for example).  This emphasizes the photometric accuracy needed to distinguish binaries from single main-sequence stars (such as stars in the region B of Figure ~\ref{fig:region}).

\subsection{The Model for the Superposition of Stars}\label{sec:blending}
The most important source of contamination in identifying binaries arises from the superposition of two unrelated stars that occurs by chance when the star cluster is projected from 3D to 2D. Two or more stars along the line of sight and within the minimum angular resolution will be measured as one star, and are indistinguishable on the CMD from real binaries. It is very difficult to screen for blended stars, but statistically, we can determine their distribution through Monte-Carlo simulations. The blending fraction is proportional to the radial stellar number density of a cluster, decreasing from the core to outside region. The general way to measure the blending fraction is through artificial star tests, in which fake stars from MSRL are added to real images and are recovered with the same photometric processes as real stars. Any fake stars overlapping with real stars will represent a blended source, and the recovered magnitudes will be their sum. As long as the added fake stars follow the radial light distribution of a globular cluster, it represents the similar blending fraction as that cluster. The drawback for this method is that it is computationally time-consuming. Consequently, we performed the Monte-Carlo simulations for an appropriate set of conditions and then modeled the results analytically, using Poisson statistics (see the Appendix).

In each simulation, we added fake stars to the real star map, where the fake stars have the same luminosity function and light distribution as the real ones, so it should have a similar blending fraction as the observed cluster. To generate the fake star list, we chose fake stars with their V magnitudes sampled by the observed luminosity function, and their I magnitudes calculated from the MSRL (i.e. all the fake stars are on the MSRL and have the same luminosity function as the real stars). The number of total fake stars added is equal to that of the total observed stars to produce a similar condition. Each fake star was assigned a radius r from the center of the cluster, which was sampled by the radial light distribution of real stars, so that the radial light distribution of fake stars is similar to the real ones. Then the azimuth angle is randomly assigned to the fake star to calculate the coordinates (x,y). To that position (x,y), any real star with distance less than 1 resolution element (the minimum resolution size Dolphot used) will be considered as a blended star, and the total magnitude is added. To find those blended stars, we used our k-dimensional tree algorithm by comparing lists of the observed stars and fake stars. To those blended sources we added photometric errors from the observed stars (see Figure ~\ref{fig:err}). We repeated this simulation 30 times to obtain an average number or an average residual color distribution  of blended stars.

\subsection{The Model for the Field Star Population} \label{sec:field star}
Field stars, both faint foreground stars and bright background stars, can also contaminate the binary population on CMDs and thus affect the accuracy of the measurement of the binary fraction, especially for low Galactic latitude clusters. They affect the number of binaries in the binary region on the CMD, as well as the number of single stars on the main-sequence. High galactic latitude clusters do not have many contaminating field stars. For example, NGC 4590 ($b=36.05^\circ$) only has 24 field stars (simulated from the model of \citet{Robin03}) in the ACS field of view, or 0.06\% of the total observed stars in that field. Low Galactic latitude clusters, however, have many contaminating field stars, such as NGC 6624 ($b=-7.91^\circ$), which has 51,325 field stars in the ACS field of view, about half of the total observed stars in that field. The best way to select cluster members is by proper motion, which, however, requires at least two epochs of HST observations separated by years. An alternative way is to construct the field star model from the theoretical model of the Galaxy to statistically account for those field stars. We used the Stellar population Synthesis model of the Galaxy \citep{Robin03} to simulate field stars at the cluster position, with the size of ACS CCD chips, $202\arcsec$  by $202\arcsec$. The V and I Johnson-Cousin magnitudes of the generated field stars are corrected for the reddening first and are converted into ACS F606W and F814W magnitudes by the transformations of \citet{sirianni05}, then are randomly added to the original images along with Poisson noise. Then we adopt the same photometry method with Dolphot to recover those field stars. The recovered field stars will have similar photometry errors as the cluster stars, as well as the completeness and blending effect.

\subsection{The Estimate of the High Mass-ratio Binary Fraction}
To estimate the high mass-ratio binary fraction, we divided the MSRL-subtracted CMD into different regions to count stars (see Figure ~\ref{fig:region}, left panel). In Figure ~\ref{fig:region}, the green line is the equal-mass binary population. The red line on the right side of the main sequence is the binary population with a binary mass-ratio of 0.5. The binary mass-ratio of 0.5 is chosen because in most cases, this binary population is beyond the 3 $\sigma$ photometric errors of the main-sequence stars.
The dashed blue lines are the upper and lower limits of usable stars for both the F606W magnitude and the residual color.
The main-sequence star (or single star) region $S$ is defined between the red lines. The binary region $B$ is where the residual color is greater than the 0.5 binary mass-ratio curve and less than 0.2 (i.e. on the right side of the red lines but on the left side of dash blue line). The residual region $R$ is where the residual colors are beyond the blue side of the symmetric line for the 0.5 binary mass-ratio curve but greater than -0.2.

We count stars separately in those regions for three types of star populations, observed stars, simulated blended stars, and simulated field stars. The simulated blended stars and field stars are generated with methods mentioned in Section ~\ref{sec:blending} and ~\ref{sec:field star}. Region S contains mainly single stars (main-sequence stars), with some contaminating field stars. Region R contains main-sequence stars with large photometric errors, and with some contaminating field stars. Region B contains mainly binaries with mass-ratios greater than 0.5, with some blended stars and field stars. It also has some main-sequence stars with large photometric errors, and can be estimated with the number in region R. So the high mass-ratio binary fraction $fb(high q)$ can be calculated by the expression:
\begin{equation}
  fb(high q)=\frac{n^{obs}_B-n^{blend}_B-n^{field}_B-(n^{obs}_R-n^{field}_R)}{n^{obs}_t-n^{field}_t}
\end{equation}
where $n^{obs}_B$, $n^{blend}_B$, and $n^{field}_B$ are the star numbers in region B for the observed stars, blended stars, and field stars, respectively.
$n^{obs}_R$ and $n^{field}_R$ are the star numbers in region R for observed stars and field stars, respectively. $(n^{obs}_R-n^{field}_R)$ represents any residual main-sequence stars with large photometric errors. This number should be also subtracted from region B assuming a symmetric distribution. The quantities $n^{obs}_t$ and $n^{field}_t$ are the total star numbers in the dashed blue line box for observed stars and field stars, which is the whole restricted region we use during the analysis. The 1 $\sigma$ error estimate for the binary fractions is estimated using Poisson errors and error propagation.

\section{The Global Binary Fractions}\label{sec:global Fb}
The high mass-ratio ($>0.5$) binary fraction discussion in Section ~\ref{sec:high-q fb} only accounts for those binaries with large deviations from the main seqeuence. Binaries with small mass ratios, however, are ignored, as they are too close to the main-sequence and are hidden by the photometric errors.  In this section, we show that, to within certain limits, one can statistically recover even small mass-ratio binaries hidden in the main sequence as long as the photometric errors are and the rate of superposition of stars are understood.


\subsection{The Star Counting Method}\label{sec:count}
To avoid contamination from photometric errors, we followed three procedures.  First, we only select stars 3 to 4 magnitudes below the turn-off point to exclude faint stars with larger photometric errors. Second, we add two gaussian models to represent the photometric errors during fitting, one for the main component with similar small errors, and the other for stars with larger errors. Tests show this to be a suitable strategy.  Third, we introduce the parameter $q_{cut}$ in the binary model, which represents the minimum binary mass-ratio that can be extracted from the data.

To estimate this, we first fit the residual color distribution with two Gaussian models, which represent the photometric errors of the main-sequence stars. The fit is only applied from 0.02 on the red side to all the blue side, since the blue side is not affected by both blending stars and real binaries, the broadening is only due to photometric errors. The fit is quite good (with $\chi^2$ close to 1). The positive residual of the fitting on the red side is due to binaries, with contamination from blending stars and field stars.

After subtracting the photometric errors, field stars, and blended stars, the residual color distribution on the red site is only from the contribution of physical binaries. We summed all the residuals on the red side and divided it by the total number of stars without the field stars, which gives the binary fraction including low mass-ratio binaries.

\subsection{The $\chi^2$ Fitting Method}\label{sec:binary}
There is another way to account for low mass-ratio binaries, and one can even constrain the binary mass-ratio distribution function. Here, we constructed the binary population by fitting to an additional binary mass-ratio distribution function.

   We assume the binary mass-ratio distribution function has the following power-law form: $f(q)\varpropto q^x$, where $q\equiv M_s/M_p$, and with a minimum value of $q_{min}$, and a maximum value of 1 (equal mass binaries). The minimum mass for the secondary star is set to 0.2 $M_{\odot}$ (the observed lowest mass from luminosity function of F606W band), thus this will set a minimum value of $q_{min}$ not to be 0.
First we assume here three different cases for the power of $x$, $x=0$, a flat mass-ratio distribution; $x=-1$, which leads to a peak at low mass ratios; $x=1$, which leads to a peak at high mass ratios. Whenever there are enough stars, we can fit for $x$ rather than assign a value.

   To construct the physical binary population, we first assume a binary fraction of $f_b$, or a total of $N*f_b$ stars to be in the binary systems, where N is the total number of the observed stars. We then extract V magnitudes with the number $N*f_b$ from the observed V magnitude luminosity function to be the V magnitudes of the primary stars.  The V magnitudes of the secondary stars are calculated using a mass-ratio q extracted from an assumed mass-ratio distribution and an assumed mass-luminosity relationship: $L \varpropto M^{3.5}$. Their I magnitudes are derived from the MSRL, and the combined binary magnitudes are calculated. Each binary system has added to the photometric errors at their V and I magnitudes to simulate the photometric spreads. Finally, we can obtain the residual color distribution for the constructed binary population by subtracting the color of the MSRL.
   We repeat this process 30 times for each binary fraction $f_b$, and construct the average profile for the physical binary population as the way we applied it to blended stars.

   Now, we have models for single stars (two gaussian models), field stars (see Section ~\ref{sec:field star}) , the superposition of stars (see Section ~\ref{sec:blending}), and binary model with the known fraction $f_b$ (see Section ~\ref{sec:binary}). The total sum of all these models will produce the final model that is compared to the observed residual color distribution profile with a $\chi^2$ test. For a given cluster, the models for single stars, field stars, and the blending stars are fixed as they depend only on the observed luminosity function, the Galactic positions, and the observed light distribution. The only model that changes during the fitting process is the binary model, as it varies with the binary fraction $f_b$, and the power law index $x$ from the binary mass-ratio distribution function.

   We developed the bisection method to search the best-fit $f_b$ (i.e. fits with the minimum chi-square value) at each assumed x. In this method, we first chose a wide initial range of binary fraction values (initial range is from 0 to 1). The chi-square values (model comparing to observed counts) were calculated for three binary fraction values, left most, right most, and the middle. Then the chi-square values at the left and right were compared to the middle one, and the left most or right most binary fraction value will be replaced by the middle one if its chi-square value is greater than the middle one. So the search range for the binary fraction now is reduced by half, and we calculated the chi-square value at the middle for the new range, and repeated the process again until the difference of chi-square values or the binary fraction range approached limiting value. The final middle value of the binary fraction range is the best-fit binary fraction with the minimum chi-square value. For clusters with enough bins, we also fit the power of x.

\begin{figure}\centering
\includegraphics[width=2.85in,angle=90 ]{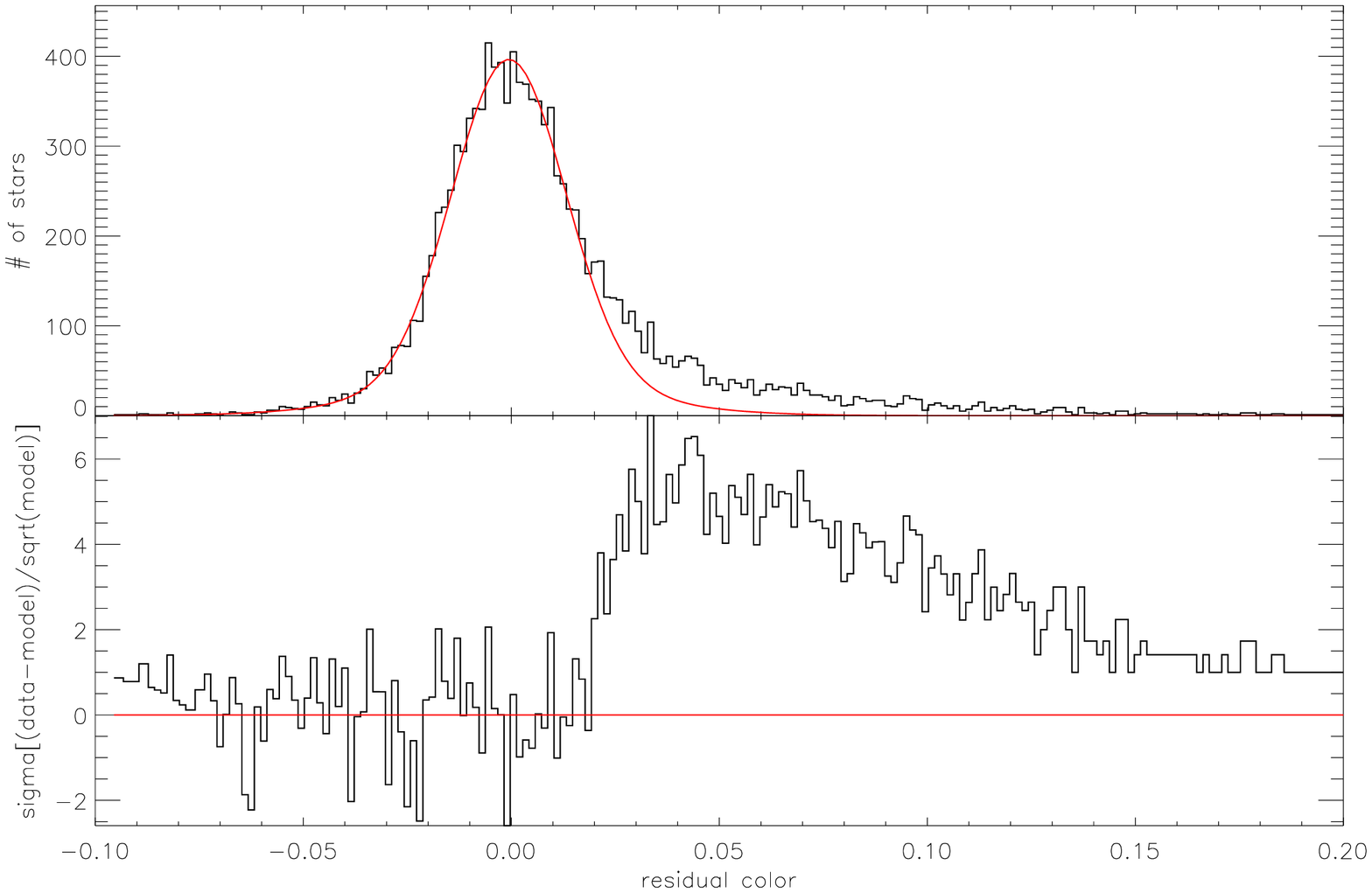}
\includegraphics[width=2.85in,angle=90 ]{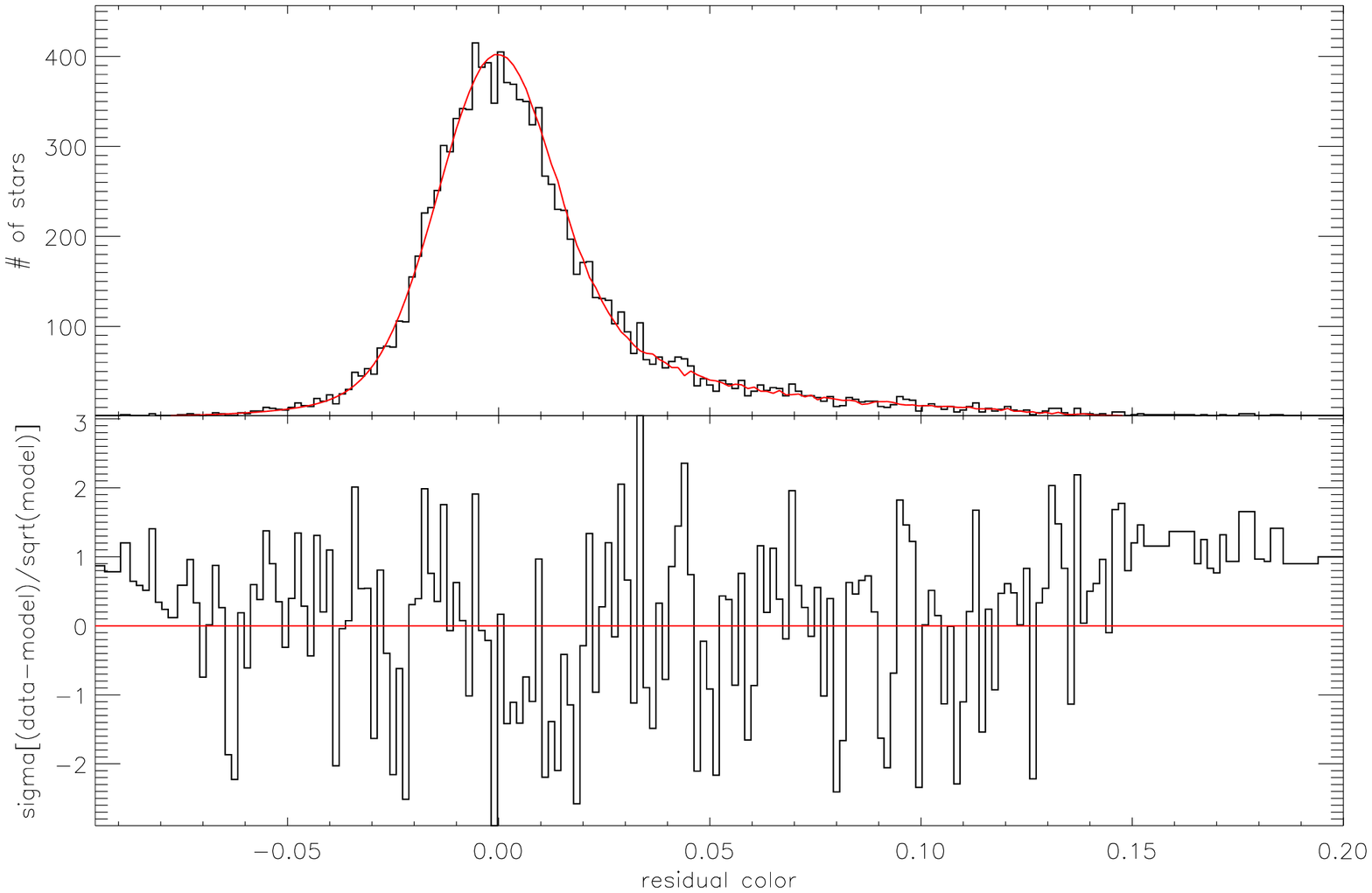}
\caption[The residual color distribution fitted by a Gaussian and by the best-fit model]{The residual color distribution fitted by a Gaussian (upper) and by the best-fit model (lower). The symmetric spread of the main-sequence is due to photometric errors, which can be fitted by Gaussian model fairly well, as there is no large systematical residuals on the blue side. The asymmetric spread of the main-sequence is due to binary populations and blending of stars, which is shown as positive residuals on the red side on the upper panel. In the lower panel, models for binaries and the blending of stars are included, and this best-fit model fits well to the observed data. }
\label{fig:ngc4590_dist}
\end{figure}

\begin{figure}\centering
\includegraphics[width=2.85in,angle=90 ]{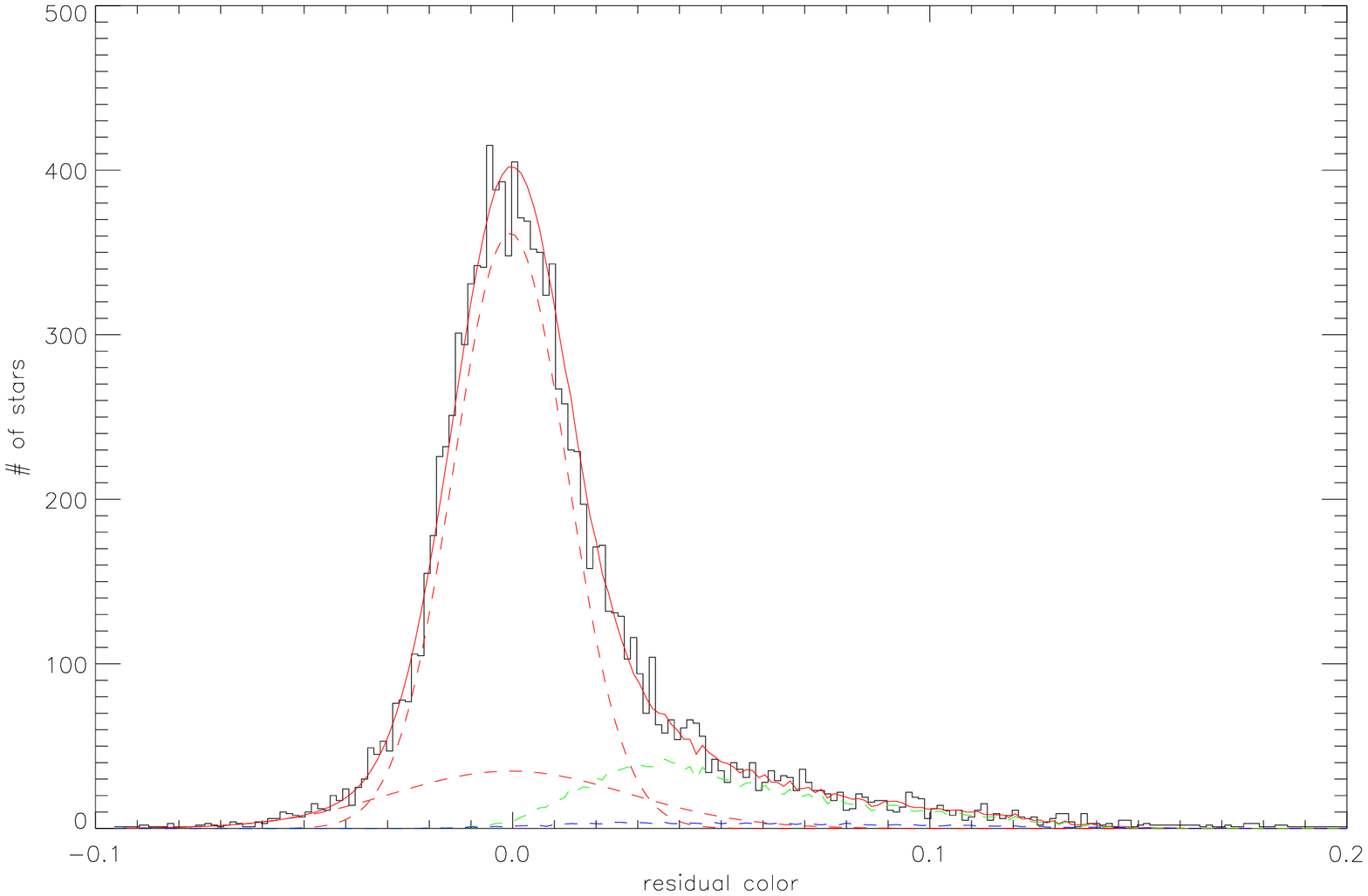}
\includegraphics[width=2.85in,angle=90 ]{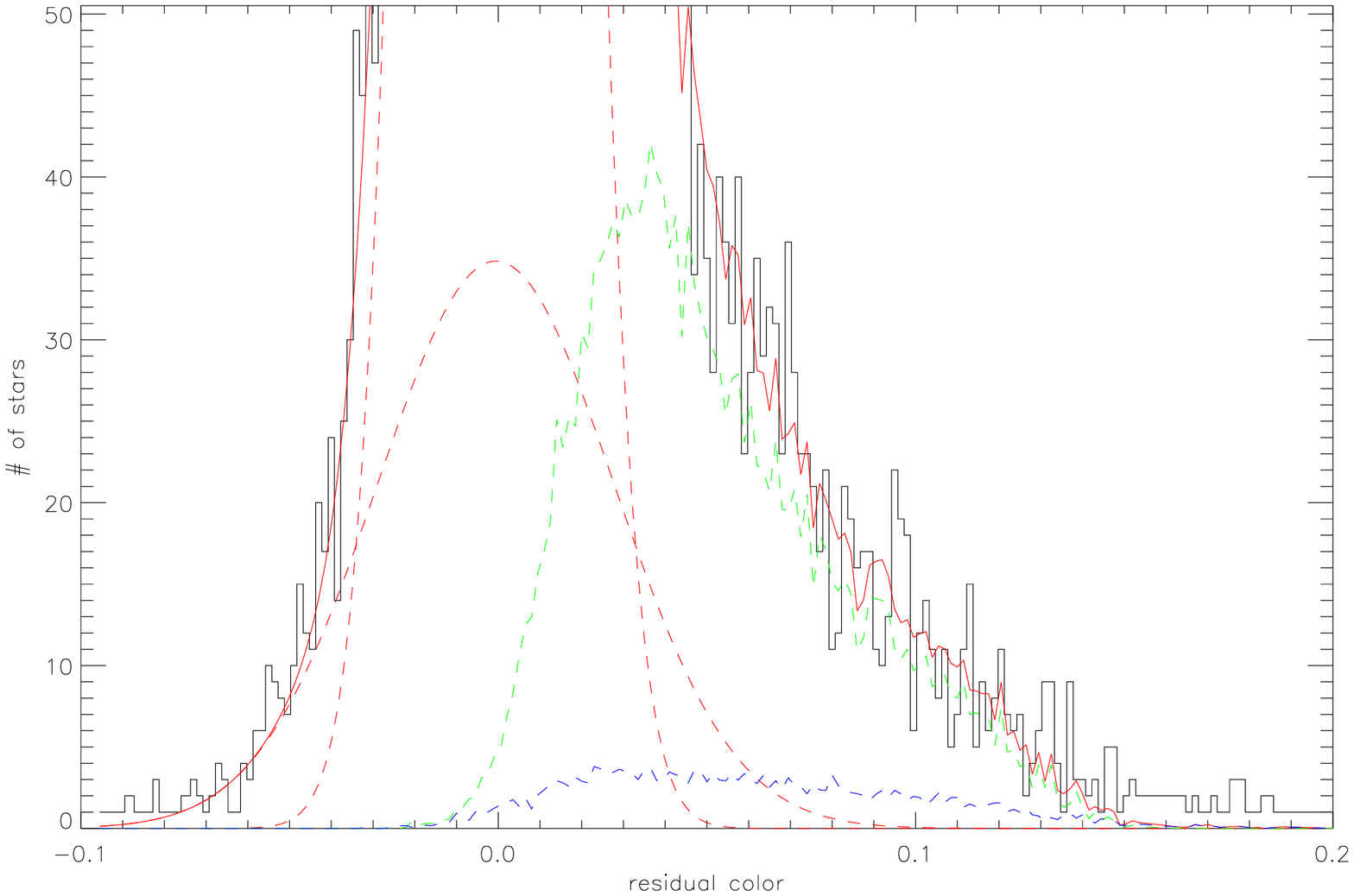}
\caption[Model components for the fitting]{Model components for the fitting (upper) and the enlarged view (lower). Solid red line: overall model; dashed red line: Gaussian model for photometric errors; dashed green line: binary model; dashed blue line: model for blending due to overlapping stars; black histogram: observed residual color distribution. For this cluster, the field star number is about 50 in the field of view of HST, so they are negligible.}
\label{fig:ngc4590_fit}
\end{figure}

   A fitting example is shown in Table ~\ref{tab:fit}, for NGC 4590. The first three rows in the table show the fitting results with the power $x$, and the minimum binary mass ratio $q_{min}$ fixed, only with the binary fraction as a free parameter in the fit. The error range in the table is estimated by changing the parameter so that the $\chi^2$ value changes by 1.0 (or 1 $\sigma$ confidence level). The best-fit value favors the model with the power $x=-1$, which also gives a higher value of binary fraction $(10.8\pm0.4)\%$. With the power $x$, and $q_{min}$ free to fit, the fit improves significantly ($\chi^2/dof=88.4/82$), with a binary fraction of $(10.8\pm0.3)\%$. Figure ~\ref{fig:ngc4590_dist} shows the residual color distribution fitted by using only the Gaussian model (upper) and with the best-fit model (lower) for NGC 4590. The symmetric spread of the main-sequence is due to photometric errors, which can be fitted by the Gaussian model fairly well, as there are no large systematic residuals on the blue side. The asymmetric spread of the main-sequence is due to binary populations and blending of stars, which is shown as positive residuals on the red side on the upper panel. In the lower panel, models for binaries and blending of stars are included, and this best-fit model fits well to the observed data.
Figure ~\ref{fig:ngc4590_fit} shows the model components for the fitting (upper) and the enlarged view (lower) for NGC 4590. For this cluster, we estimate about 50 field stars in the field of view of HST, so they are negligible contaminant. In Table ~\ref{tab:method}, we show the binary fractions estimated with the three methods discussed above, the high mass-ratio method (the second column), the star counting method (the third column), and the $\chi^2$ fitting method (the fourth column), and we expect the binary fraction obtained with the global methods (the star counting method and the $\chi^2$ fitting method) to be higher than the high mass-ratio method, as we approach lower mass-ratio values.

\section{The Binary Fraction Radial Analysis} \label{sec:radial}
One prominent predicted effect of globular cluster dynamical evolution is mass segregation, which implies that massive stars tend to sink to the center of the core while light stars are redistributed to the outside of the cluster. Since a binary system contains two stars, it is more massive than a single star, so they tend to sink towards the cluster core by mass segregation. By performing radial analysis of binary fractions, we can test for this dynamical effect.

\subsection{The Analysis Method}
In this analysis, we divide the whole ACS field of view into three annular bins, with their centers at the cluster center obtained from \citet{harris96}. The bin sizes were chosen (iteratively) so that each bin contains roughly one-third of the total stars recovered from the whole field, which leads to similar error bars for the binary fraction in each region.

\subsection{The Radial CMD Qualities and the Example of the Results}
For high density clusters, the CMD quality for the central bin is the worst among the three, which shows larger photometric spread and a lower faint star recovery rate than the other two. This is understandable, as the higher star not only increases the background level, making fainter stars more difficult to detect, but also increases the blending probability, making the PSF determination poorer.  For low density clusters, we do not observe large variations in the CMD qualities.
							
\begin{figure}\centering
\includegraphics[width=4in,angle=0]{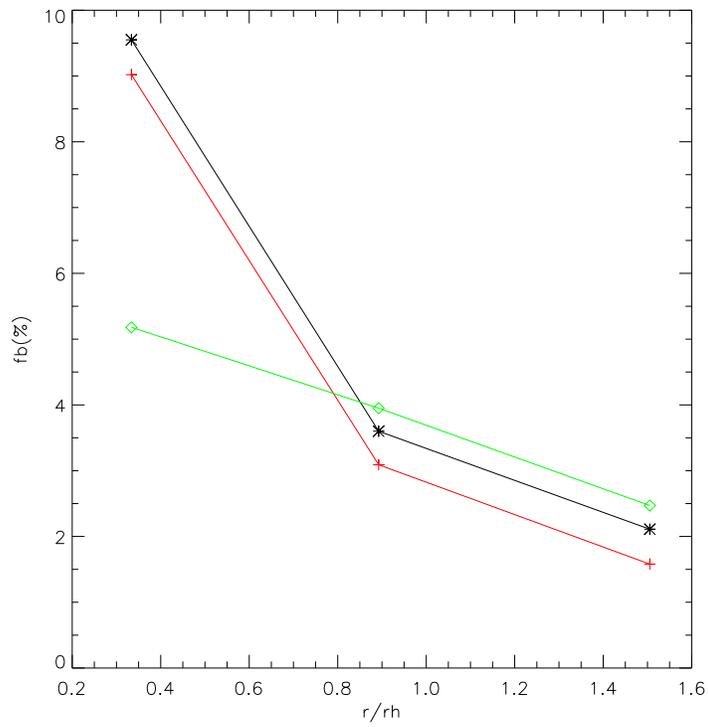}
\caption{Binary fraction radial distribution for NGC 6981.}
\label{fig:radial}
\end{figure}

In Table \ref{tab:radial} and Fig. \ref{fig:radial}, we show the results of the radial analysis on the binary fraction for NGC 6981 as an example. In Table \ref{tab:radial}, we list the sizes of the annular bins, binary fractions obtained by the high mass-ratio method, the counting method, and the $\chi^2$ fitting method, and the dof/$\chi^2$ for the fitting method. In Fig. \ref{fig:radial}, we plot the binary fractions obtained from different methods and from different annular bins against their positions relative to the cluster center, in units of the half mass radius. We clearly see a decreasing trend of the binary fraction towards the outside of this cluster, an effect discussed for the sample of 35 globular clusters in Paper II.

\section{Additional Factors Affecting the Binary Fraction and Final Comments}\label{sec:discussion}
In this section, we will discuss the additional factors that can affect the measurement accuracy of binary fractions in globular clusters using their CMDs.

\subsection{The Photometric Errors}
The key aspect to estimating the binary fraction using the CMD method is the highly accurate photometry for the CMDs. As the photometric errors become larger, the spread of the main-sequence becomes larger, which will smear out the signals from binaries with small mass-ratio. In order to decrease the photometric errors, we need to increase the exposure time. This is true for low density clusters, in which stars are quite isolated, and the PSF can be determined quite well. For most low density clusters in our sample, the photometric errors are approaching their theoretical limit. High density clusters, however, have very crowded central regions, where the high and varied background make the PSF determination uncertain. Most of the errors are not from low S/N ratios but from uncertainties in the PSFs and the backgrounds. Increasing the observing time for those crowded clusters would not help lower the photometric errors. Instead, developing more sophisticated PSF photometry algorithm for crowded region is needed, which is still not fully mature.

\subsection{The Metallicity Dispersion}
\begin{figure}\centering
\includegraphics[width=4in,angle=0]{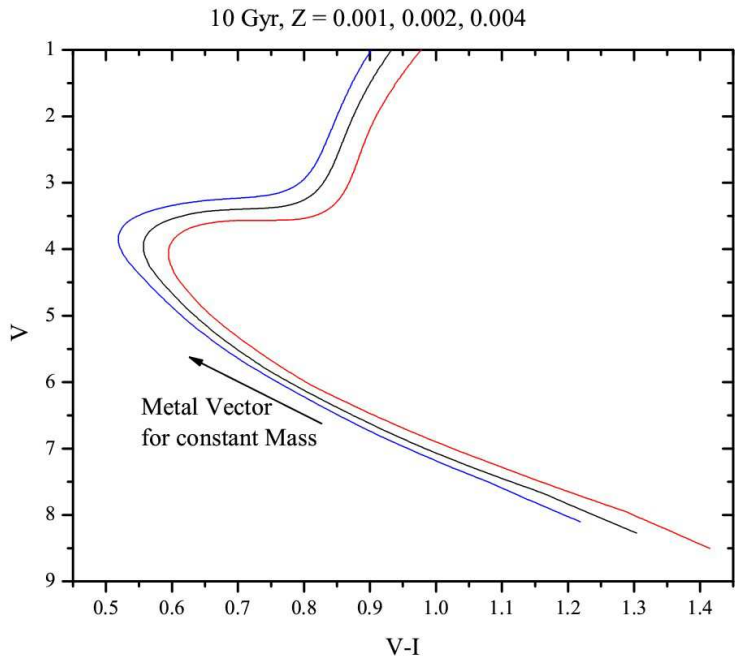}
\caption{Isochrone Models at different metallicity.}
\label{fig:metal}
\end{figure}

Theoretical modeling shows that the dispersion of the metallicity of a globular cluster can also cause a spread in color on the main sequence. Figure ~\ref{fig:metal} shows that a metallicity ($Z$) difference of 0.002 (from red to blue line at certain V magnitude), equivalent to $\delta[Fe/H]=0.30$, can cause a spread of about 0.054 magnitude in color, and 0.284 in the V magnitude. So given the uncertainty in [Fe/H] of 0.03, the spread in color will be 0.005, and shift in V magnitude will be 0.028. The observed intrinsic color spread is smaller compared to the typical width in color of the main-sequence with the HST observations (about 0.012 for NGC 5053 in our sample), and the shift in V magnitude will not affect the color distribution. Thus the intrinsic metallicity dispersion can be negligible. For multi-population systems (such as NGC 2808), however, this will not be the case \citep{Piotto07,Pasquini11}.

\subsection{The Differential Reddening}
Observations on globular clusters located near the Galactic center can be affected by the existence of large and differential extinction of the foreground dust. \citet{alonso-Garcia11} discuss a technique to correct this effect. From our sample, most clusters are well above the Galactic plane and with the reddening E(B-V) less than 0.1, which are not important to spread the color of the main sequence comparing to their photometric errors. For clusters near the Galactic plane, however, the reddening can be very large. Along with the heavily contaminated field stars, the determination of binary fractions in those clusters are quite uncertain.

\subsection{Comparison with Another Survey}
At the same time as this work was being carried out, another group was working toward a similar goal and recently published their comprehensive work \citep{milone12}.  Although both efforts follow established approaches that use the CMD, there are some differences, which we identify and compare the values derived from the two independent approaches.  One important difference is the software used to obtain photometry in crowded fields, which makes extensive use of the psf libraries.  We used DOLPHOT (V1.2) while \citet{milone12} used a the proprietary algorithms described by \citet{anderson08}, developed specifically for crowded field photometry.  We used the stellar field model of \citet{Robin03} while \citet{milone12} used the model of \citet {girardi05}.
	
Both we and \citet{milone12} performed extensive artificial star tests although with slight differences in how completeness was defined and how finely the globular cluster stellar density was subdivided.  Most other procedures were essentially identical, including spline ridge-line fitting to define the Main Sequence or the magnitude range of the Main Sequenced used for analysis.  The two clusters described here have small reddening, so we did not need the sophisticated corrections applied by \citep{milone12}, nor were there multiple epochs of data to be considered.

The two clusters that we discuss here were also analyzed by \citet{milone12} and we find general agreement, although the results are expressed slightly differently.
For NGC 4590, our binary fractions for $q>0.5$ and within the half mass radius was 6.2 $\pm$ 0.3 \% while  \citep{milone12} obtain a somewhat lower value of 5.3 $\pm$ 0.7 \%.  For the total binary fraction, \citep{milone12} doubles the $f(q>0.5)$ value, which assumes a flat distribution in $q$, obtaining 10.6 $\pm$ 1.4 \%.  We fit a flat functional form to the CMD distribution and obtain a binary fraction of 9.4 $\pm$ 0.7 \%.
Since these are the same data sets, the differences, which are comparable to the uncertainties, most likely reflects systematic differences between the approaches.

The other comparison that can be made is the radial distribution of NGC 6981, which we find drops by about a factor of 4-5 from a bin within $r_h$ to one that extends to 1.9$r_h$ (from 9.6 $\pm$ 0.6\% to 2.1 $\pm$ 0.3\%).  This decrease is similar to Milone's mean result for their sample \citep{milone12} but for this particular object, they find a smaller decline, from about 5\% to 3\%, with error bars of about 1\% for each value.  The reason for this difference is not obvious to us.  In a companion paper, we will compare our sample to theirs, which will provide the statistical power to identify significant differences and systematic effects.

\section{Final Comments and Summary}
Binary stars are thought to be a controlling factor in globular cluster evolution. To systematically study them, we conducted this survey of 35 Galactic globular clusters, taking advantage of the wealth of the HST data. In this paper, the first of two, we present the techniques used in obtaining their binary fractions. We used the PSF-fitting photometry with DOLPHOT (V1.2) to obtain high quality color-magnitude diagrams. We applied three different methods to estimate the binary fractions. The high mass-ratio method, a model-independent method, counts the number of binaries extending above a binary mass-ratio of 0.5 on the color-magnitude diagram. The star counting method also takes into account the low mass-ratio binaries after modeling the main-sequence population, star superposition, and the field stars. The $\chi^2$ fitting method not only estimates the binary fraction, but also models the binary mass-ratio distribution. We showed a representative globular cluster NGC 4590, with a constrained binary fraction in the range of 6.2\% to 10.8\% by the three methods. To test the effect of globular cluster dynamical evolution, we introduced the binary fraction radial analysis with NGC 6981 as an example, which shows a decreasing trend of binary fraction towards the outside of this cluster. We also discussed the factors that could affect the accuracy of measuring the binary fraction with our methods.

In Paper II, we will show the results of this survey, including accurate color-magnitude diagrams, the binary fractions within the core and the half mass radius obtained with three methods, the radial binary fraction analysis, and the potential binary candidate list for further observation. We will compare our observational results to the theoretical predictions of the globular cluster dynamical evolution.

\section{Acknowledgements}
The authors would like to thank A.E. Dolphin for answering our many questions that arose when using the photometry package Dolphot V 1.2.
We appreciate the many thoughtful suggestions from the referee, as well as from Mario Mateo, Jon Miller, Eric Bell, Sally Oey, and Patrick Seitzer.  We gratefully acknowledge financial support through a HST grant from NASA.


{}

\centerline{\textbf{Appendix}}

The Superposition of stars is a contamination that has the same effect as real binaries on CMD. It is very difficult to screen for them, but statistically, we can estimate the number of blended stars. In this appendix, we will discuss the probability for different types of blended stars (i.e. unresolved doubles, triples, etc.) in globular clusters, which can provide a good estimate on the blending frequency for globular clusters at different stellar density.

\textbf{1) Poisson Distribution}

 The probability that one star is blended with others depends on the projected 2D star number density as well as the angular resolution. It is a Poisson process and the probability can be described by the Poisson probability distribution function:
\begin{displaymath}
  P(x)=\frac{\mu^x e^{-\mu}}{x!}
\end{displaymath}
where $x$ is the companion number for the blends, i.e. $x=1$ is for unresolved double stars (star with one companion), and $x=2$ is for unresolved triple stars (star with two companions), etc. $\mu$ is the area ratio of the minimum resolved area to the mean occupied area per star in the reference frame.

\textbf{2) Monte Carlo Blending Test}

\begin{figure}\centering
\includegraphics[width=2.5in,angle=0 ]{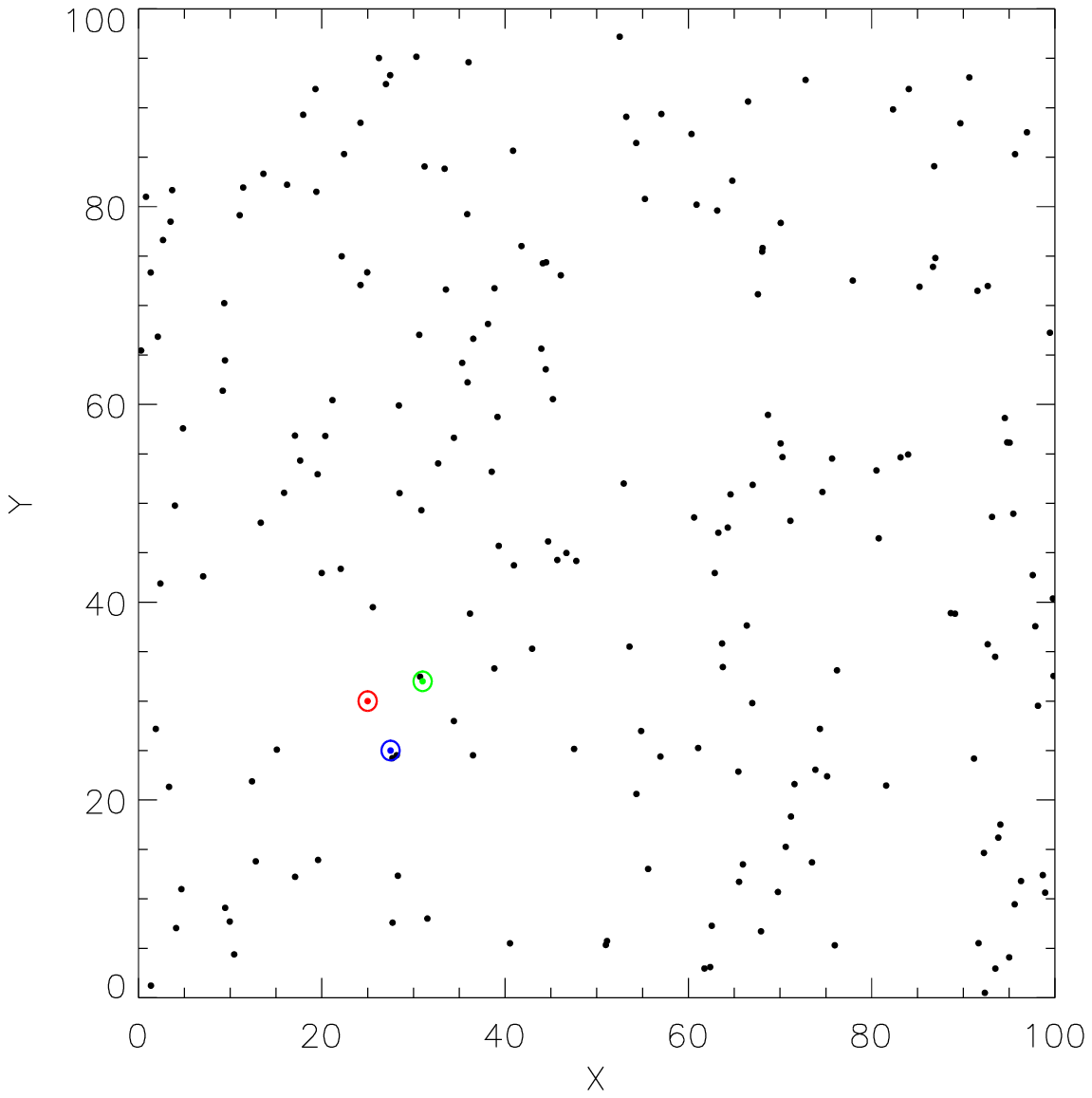}
\includegraphics[width=2.5in,angle=0 ]{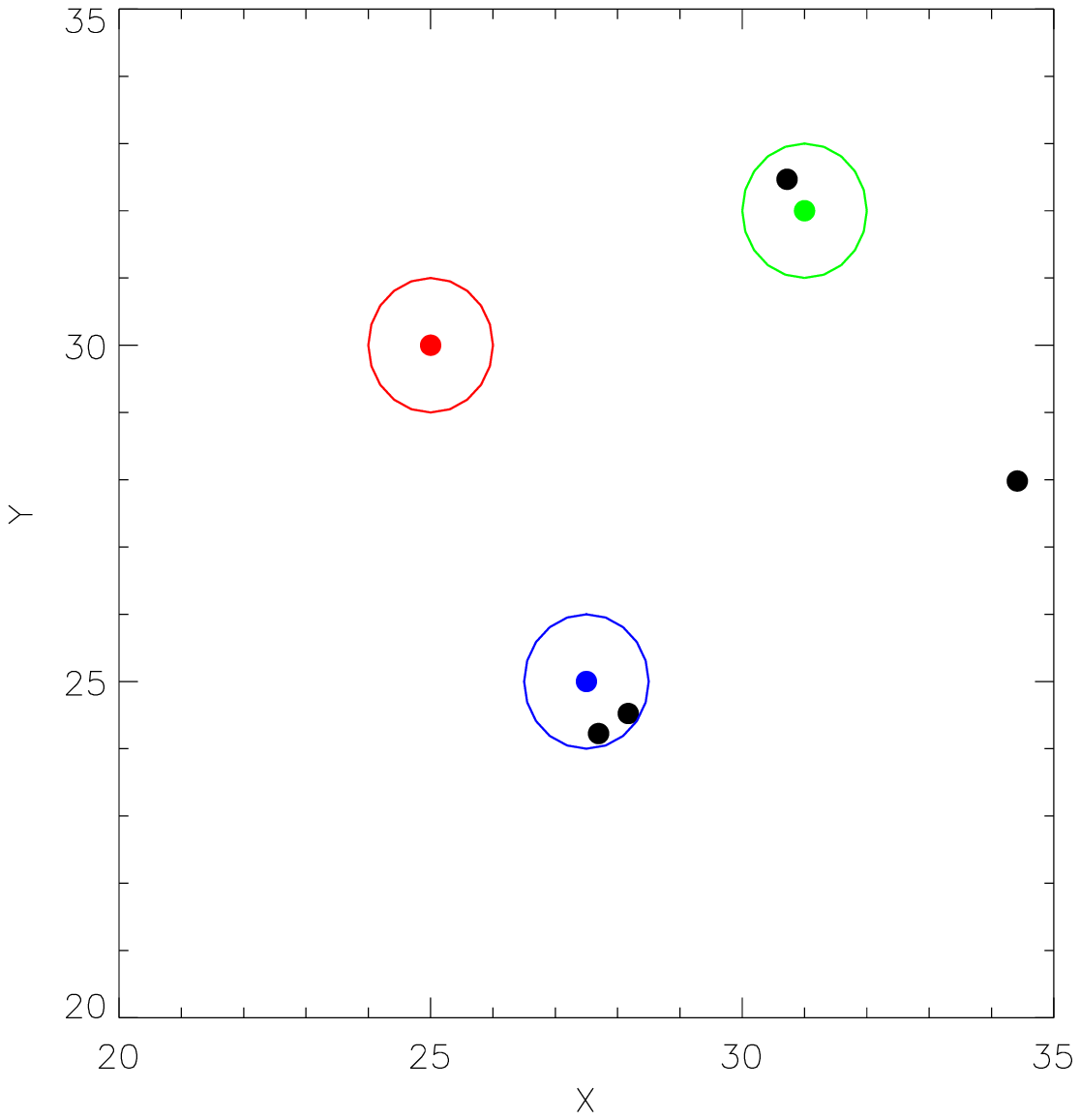}
\caption[Monte Carlo simulation rules for determining the frequency of blending of stars]{Monte Carlo simulation rules for determining the frequency of blending of stars. Left: the black dots are randomly input reference stars ($N_{total}=200$), while the color dots are randomly added test stars with a minimum resolved area within the circles. Right: an enlarged plot for the test stars. Red: added test star without a companion (not a blend); Green: added test star with one companion; Blue: added test star with two companions.}
\label{fig:setup}
\end{figure}

To test the hypothesis that the blending probability can be described by a Poisson distribution function, we performed the following Monte Carlo simulations. We randomly distributed $N_{total}$ stars in a fixed square area with each size of 100 pixels to form the reference frame. Secondly, we randomly added one test star to this reference frame. Then we counted how many reference stars are within the minimum resolution radius $r_{min}$ of the added test star, where $r_{min}=1$ to simulate the ACS image. If the count is more than 0, then the added test star will be considered as a blend. After counting, the test star was removed, and a new test star was randomly added to follow the above process. We added 5000 test stars in all for each simulation, and the final blending fraction is the total number of the blending stars divided by the total number (5000) of added test stars. Blends with different companion number were counted separately. We repeated this simulation 30 times (i.e. 30 different random distributions of the reference stars) for each star number density (i.e. each $N_{total}$) to get the mean blending fraction and the standard deviation. The simulation setup is shown in Figure ~\ref{fig:setup}.

\begin{figure}\centering
\includegraphics[width=3in,angle=0 ]{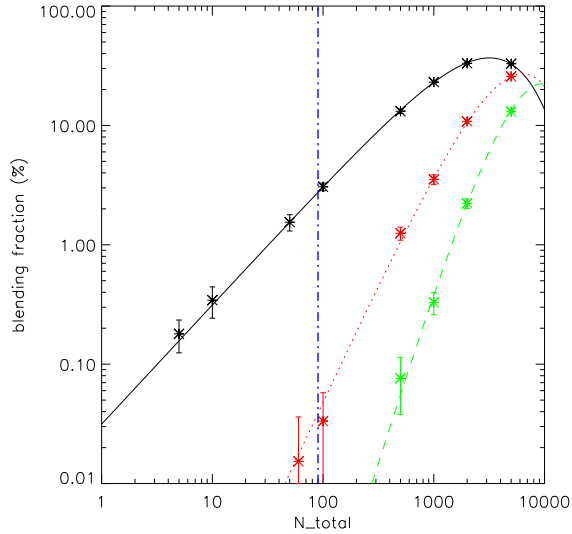}
\caption[Monte Carlo simulation results for the blending effect]{Monte Carlo simulation results for the blending effect. Each data point is the mean of 30 MC simulations with the standard deviation as the error bars. The underlying lines are the predicted Poisson distribution. Black solid line: blending fraction for blends with one companion; Red dotted line: blending fraction for blends with two companions; Green dashed line: blending fraction for blends with three companions. The blue vertical dot-dash line is where the maximum star number density cluster is in our cluster sample. The error bars for the data points with large $N_{total}$ are much smaller than the symbol size. }
\label{fig:MC_result}
\end{figure}

Results were compared to the Poisson probability distribution function (see Figure ~\ref{fig:MC_result}), where $\mu$ is defined as
\begin{displaymath}
\mu=\frac{A_{min}}{A_{mean}}=\frac{\pi*r_{min}^2}{L^2/N_{total}}=\frac{\pi*r_{min}^2}{L^2}*N_{total}.
\end{displaymath}
So for the fixed area with size L and the minimum resolution radius $r_{min}$, $\mu$ only depends on the input star number $N_{total}$.

Here we calculated the blending probability for three different blending stars, $x=1$ (unresolved double stars), $x=2$ (unresolved triple stars), and $x=3$ (unresolved quadruple stars), and compared to the Monte Carlo simulations (Figure ~\ref{fig:MC_result}).  In Figure ~\ref{fig:MC_result}, we can see that the Poisson distribution curves match those MC data points fairly well. This indicates that as long as we know the projected star number density and the minimum resolution, we can estimate the blending fraction using the Poisson distribution function.

For example, in our globular cluster sample, the maximum star number density is from NGC 7078, where we obtain 156,080 stars down to the 26th magnitude in the HST ACS CCD with the size of 4096 by 4096 pixels, which is equivalent to around 90 stars in the 100 by 100 pixel area of our Monte Carlo simulation setup (see the blue vertical dot-dash line in Figure ~\ref{fig:MC_result}). Even for this densest cluster, the blending fraction with one companion is less than 3\%, and the blending fraction with two companions is much smaller, less than 0.04\%. The typical star number recovered in our cluster sample is around 30,000 stars, which is equivalent to around 17 stars in the 100 by 100 pixel area. The blending fraction is less than 0.7\% for blends with one companion, which is only a small fraction in the total binary fraction budget. The blending fraction with a higher number of companions is two orders of magnitude smaller.

Note that here we used the average star number density to calculate the blending fraction for the whole field, which is not appropriate, as in globular clusters the star number density varies quickly along the radius. But as long as we only consider small range of radius, the density gradient will be small, and one can use this method to estimate the blending fraction in that area. The blending model in the fitting process, however, takes into account the stellar number density gradient, as it follows the observed stellar radial distribution.

\clearpage

\begin{table}
	
 \caption{Basic properties of the Galactic globular clusters in the sample$^a$}
\resizebox{5in}{!}{
		\begin{tabular}{cccccccccccc}
\hline
\hline
Object & l & b & [Fe/H]& E(B-V)& $M_V$& core & $r_c$ & $r_h$ & log $t_{rh}$ & ${t_{Gyr}}^b$ & $t_{Gyr}/t_{rh}$ \\
       & $^\circ$ & $^\circ$ &  &  &   &collapse &$^\prime$  &$^\prime$ & &Gyr& \\
\hline
 NGC 104& 305.89& -44.89& -0.72& 0.04&  -9.42&    & 0.36& 3.17&  9.55&  10.7&   3.0\\
  NGC 288& 152.30& -89.38& -1.32& 0.03&  -6.75&    & 1.35& 2.23&  9.32&  11.3&   5.4\\
  NGC 362& 301.53& -46.25& -1.26& 0.05&  -8.43&  y& 0.18& 0.82&  8.93&   8.7&  10.2\\
 NGC 1851& 244.51& -35.03& -1.18& 0.02&  -8.33&    & 0.09& 0.51&  8.82&   9.2&  13.9\\
 NGC 2808& 282.19& -11.25& -1.14& 0.22&  -9.39&    & 0.25& 0.80&  9.15&   9.3&   6.6\\
 NGC 4590& 299.63&  36.05& -2.23& 0.05&  -7.37&    & 0.58& 1.51&  9.27&  11.2&   6.0\\
 NGC 5053& 335.70&  78.95& -2.27& 0.01&  -6.76&    & 2.08& 2.61&  9.87&  10.8&   1.5\\
      M 3&  42.22&  78.71& -1.50& 0.01&  -8.88&    & 0.37& 2.31&  9.79&  11.3&   1.8\\
 NGC 5466&  42.15&  73.59& -1.98& 0.00&  -6.98&    & 1.43& 2.30&  9.76&  12.2&   2.1\\
 NGC 5897& 342.95&  30.29& -1.90& 0.09&  -7.23&    & 1.40& 2.06&  9.57&  12.3&   3.3\\
 NGC 5904&   3.86&  46.80& -1.29& 0.03&  -8.81&    & 0.44& 1.77&  9.41&  10.9&   4.2\\
 NGC 5927& 326.60&   4.86& -0.49& 0.45&  -7.81&    & 0.42& 1.10&  8.94&   10.9$^c$& 12.5 \\
 NGC 6093& 352.67&  19.46& -1.75& 0.18&  -8.23&    & 0.15& 0.61&  8.80&  12.4&  19.7\\
 NGC 6121& 350.97&  15.97& -1.16& 0.35&  -7.19&    & 1.16& 4.33&  8.93&  11.7&  13.7\\
 NGC 6101& 317.74& -15.82& -1.98& 0.05&  -6.94&    & 0.97& 1.05&  9.22&  10.7&   6.4\\
     M 13&  59.01&  40.91& -1.53& 0.02&  -8.55&    & 0.62& 1.69&  9.30&  11.9&   6.0\\
 NGC 6218&  15.72&  26.31& -1.37& 0.19&  -7.31&    & 0.79& 1.77&  8.87&  12.5&  16.9\\
 NGC 6341&  68.34&  34.86& -2.31& 0.02&  -8.21&    & 0.26& 1.02&  9.02&  12.3&  11.7\\
 NGC 6352& 341.42&  -7.17& -0.64& 0.22&  -6.47&    & 0.83& 2.05&  8.92&   9.9&  11.9\\
 NGC 6362& 325.55& -17.57& -0.99& 0.09&  -6.95&    & 1.13& 2.05&  9.20&  11.0&   6.9\\
 NGC 6397& 338.17& -11.96& -2.02& 0.18&  -6.64&   y& 0.05& 2.90&  8.60&  12.1&  30.4\\
 NGC 6541& 349.29& -11.19& -1.81& 0.14&  -8.52&  y& 0.18& 1.06&  9.03&  14.0$^d$&  13.1\\
 NGC 6624&   2.79&  -7.91& -0.44& 0.28&  -7.49&   y& 0.06& 0.82&  8.71&  10.6&  20.7\\
 NGC 6637&   1.72& -10.27& -0.64& 0.18&  -7.64&    & 0.33& 0.84&  8.82&  10.6&  16.0\\
 NGC 6652&   1.53& -11.38& -0.81& 0.09&  -6.66&    & 0.10& 0.48&  8.39&  11.4&  46.4\\
 NGC 6656&   9.89&  -7.55& -1.70& 0.34&  -8.50&    & 1.33& 3.36&  9.23&  12.3&   7.2\\
 NGC 6723&   0.07& -17.30& -1.10& 0.05&  -7.83&  y& 0.83& 1.53&  9.24&  11.6&   6.7\\
 NGC 6752& 336.49& -25.63& -1.54& 0.04&  -7.73&   y& 0.17& 1.91&  8.87&  12.2&  16.5\\
 Terzan 7&   3.39& -20.07& -0.32& 0.07&  -5.01&    & 0.49& 0.77&  8.96&   7.4&   8.1\\
    Arp 2&   8.55& -20.79& -1.75& 0.10&  -5.29&    & 1.19& 1.77&  9.70&  11.3&   2.3\\
 NGC 6809&   8.79& -23.27& -1.94& 0.08&  -7.57&    & 1.80& 2.83&  9.29&  12.3&   6.3\\
 NGC 6981&  35.16& -32.68& -1.42& 0.05&  -7.04&    & 0.46& 0.93&  9.23&   9.5$^e$&   5.6\\
 NGC 7078&  65.01& -27.31& -2.37& 0.10&  -9.19&   y& 0.14& 1.00&  9.32&  11.7&   5.6\\
 NGC 7099&  27.18& -46.84& -2.27& 0.03&  -7.45&   y& 0.06& 1.03&  8.88&  11.9&  15.7\\
Palomar 12&  30.51& -47.68& -0.85& 0.02&  -4.47&    & 0.02& 1.72&  9.28&   6.4&   3.4\\
\hline
\end{tabular}
}

\flushleft
$^a$ all columns except ages $t_{Gyr}$ are from \citet{harris96}.\\
$^b$ ages $t_{Gyr}$ are from \citet{salaris02} except $^c$ from \citet{fullton96}, $^d$ from \citet{Alonso-Garcia10}, and $^e$ from \citet{sollima07}.
\label{table:basic}
\end{table}

\begin{table}[tbp]
	\caption{HST ACS observation log}
\resizebox{3.5in}{!}{
		\begin{tabular}{ccccc}
\hline
\hline
Name& Filter  & \# of Exposures & Exposure Time (s) & Dataset\\
\hline
NGC 104 & F606W & 4 &50 &J9L960010       \\
        & F814W & 4 &50 &J9L960020       \\
NGC 288 & F606W & 4 &130 &J9L9AD010       \\
        & F814W & 4 &150 &J9L9AD020       \\
NGC 362 & F606W & 4 &150 &J9L930010       \\
        & F814W & 4 &170 &J9L930020       \\
NGC 1851 & F606W & 5 & 350&J9L910010       \\
        & F814W & 5 & 350&J9L910020       \\
NGC 2808 & F606W & 5 & 360&J9L947010       \\
        & F814W & 5 &370 &J9L947020       \\
NGC 4590 & F606W & 4 & 130&J9L932010       \\
        & F814W & 4 &150 &J9L932020       \\
NGC 5053 & F606W & 5 &340 &J9L902010       \\
        & F814W & 5 &350 &J9L902020       \\
M 3      & F606W & 4 &130 &J9L953010       \\
        & F814W & 4 &150 &J9L953020       \\
NGC 5466 & F606W & 5 & 340&J9L903010       \\
        & F814W & 5 & 350&J9L903020       \\
NGC 5897 & F606W & 4 & 340&J9L913010       \\
        & F814W & 3 &350 &J9L913020       \\
NGC 5904 & F606W & 4 &140 &J9L956010       \\
        & F814W & 4 &140 &J9L956020       \\
NGC 5927 & F606W & 5 & 350&J9L914010       \\
        & F814W & 5 & 360&J9L914020       \\
NGC 6093 & F606W & 5 &340 &J9L916010       \\
        & F814W & 5 &340 &J9L916020       \\
NGC 6101 & F606W & 5 &370 &J9L917010       \\
        & F814W & 5 & 380&J9L917020       \\
NGC 6121 & F606W & 4 &25 &J9L964010       \\
        & F814W & 4 &30 &J9L964020       \\
M 13  & F606W & 4 &140 &J9L957010       \\
        & F814W & 4 &140 &J9L957020       \\
NGC 6218  & F606W & 4 & 90&J9L944010       \\
        & F814W & 4 &90 &J9L944020       \\
NGC 6341 & F606W & 4 &140 &J9L958010       \\
        & F814W & 4 &150 &J9L958020       \\
NGC 6352 & F606W & 4 & 140&J9L959010       \\
        & F814W & 4 & 150&J9L959020       \\
\hline
 & & \multicolumn{3}{r}{continued on next page} \\
 \end{tabular}
 }
\label{table:obs log}
\end{table}

\begin{table}[tbp]
	  \resizebox{3.5in}{!}{
  \begin{tabular}{ccccc}
\multicolumn{3}{l}{continued from previous page}  & & \\
\hline
\hline
Name& Filter  & \# of Exposures & Exposure Time (s) & Dataset\\
\hline
NGC 6362 & F606W & 4 &130 &J9L934010       \\
        & F814W & 4 & 150&J9L934020       \\
NGC 6397 & F606W & 4 &15 &J9L965010       \\
        & F814W & 4 & 15&J9L965020       \\
NGC 6541 & F606W & 4 &140 &J9L936010       \\
        & F814W & 4 &150 &J9L936020       \\
NGC 6624 & F606W & 5 &350 &J9L922010       \\
        & F814W & 5 & 350&J9L922020       \\
NGC 6637 & F606W & 5 & 340&J9L937010       \\
        & F814W & 5 & 340&J9L937020       \\
NGC 6652 & F606W & 5 &340 &J9L938010       \\
        & F814W & 5 &340 &J9L938020       \\
NGC 6656 & F606W & 4 & 55&J9L948010       \\
        & F814W & 4 & 65&J9L948020       \\
NGC 6723 & F606W & 4 &140 &J9L941010       \\
        & F814W & 4 &150 &J9L941020       \\
NGC 6752 & F606W & 4 & 35&J9L966010       \\
        & F814W & 4 & 40&J9L966020       \\
NGC 6809 & F606W & 4 &70 &J9L963010       \\
        & F814W & 4 &80 &J9L963020       \\
NGC 6981 & F606W & 4 & 130&J9L942010       \\
        & F814W & 4 & 150&J9L942020       \\
NGC 7078 & F606W & 4 &130 &J9L954010       \\
        & F814W & 4 & 150&J9L954020       \\
NGC 7099 & F606W & 4 & 140&J9L955010       \\
        & F814W & 4 &140 &J9L955020       \\
Arp2  & F606W & 5 & 345&J9L925010       \\
        & F814W & 5 & 345&J9L925020       \\
Parlomar 12  & F606W & 5 &340 &J9L928010       \\
        & F814W & 5 & 340&J9L928020       \\
Terzan 7 & F606W & 5 &345 &J9L924010\\
        & F814W & 5 &345 &J9L924020\\
\hline
\end{tabular}
}
\end{table}

\begin{table}[htbp]
	\centering
	\caption{Comparing of the fitting results among different binary mass-ratio models for NGC 4590 using the $\chi^2$ fitting method.}
		\begin{tabular}{lccccc}
\hline
\hline
Method  & Power $x$ & $q_{min}$ & $f_b$(\%) & $\chi^2/dof$ \\
\hline
fixed $x,q_{min}$ & 0 & 0.24 &$9.4 \pm 0.7$ &175.9/84  \\
fixed $x,q_{min}$ & 1 & 0.24 & $6.8 \pm 0.2$ & 236.3/84 \\
fixed $x,q_{min}$ & -1 & 0.24 & $10.8 \pm 0.4$ &126.7/84 \\
fit all & $-2.3 \pm 1.5$ &$0.325 \pm 0.007$ & $10.3 \pm  0.4$&88.4/82\\
\hline
\end{tabular}
	\label{tab:fit}
\end{table}

\begin{table}[htbp]
	\centering
	\caption{Binary fractions within the half mass radius region with different analyzing methods for NGC 4590.}
		\begin{tabular}{lccc}
\hline
\hline
Source & $f_b(q>0.5)\%$ & $f_b(count)\%$ & $f_b(fit)\%$ \\
\hline
 ngc4590 &    6.24$\pm$ 0.30 &    7.33$\pm$1.39 &    8.62$\pm$1.21 \\
\hline
\end{tabular}
	\label{tab:method}
\end{table}

\begin{table}[htbp]
	\centering
	\caption{Example of binary fraction radial analysis for NGC 6981.}
		\begin{tabular}{cccccc}
\hline
\hline
Source & Bin Range$(r_h)$ & $f_b(q>0.5)\%$ & $f_b(count)\%$ & $f_b(fit)\%$ &  $\chi^2$/dof\\
\hline
  ngc6981 &  0.00-0.67 &    9.55$\pm$0.58 &    8.62$\pm$2.18 &    5.18$\pm$0.77 &    54.4/ 75 \\
 &  0.67-1.12 &    3.60$\pm$0.40 &    3.55$\pm$2.20 &    3.95$\pm$3.95 &    41.2/ 66 \\
 &  1.12-1.89 &    2.11$\pm$0.32 &    1.14$\pm$2.22 &    2.47$\pm$1.48 &    53.7/ 58 \\
\hline
\end{tabular}
	\label{tab:radial}
\end{table}

\clearpage

\end{document}